\documentclass[pre,aps,
preprint,%
showpacs,superscriptaddress,
amsmath,floatfix]{revtex4-2}
\usepackage{graphicx}
\usepackage[utf8]{inputenc}
\usepackage{amsmath,amssymb}
\usepackage{graphicx}
\usepackage{comment}
\usepackage{setspace}
\usepackage{CJK}
\usepackage{dcolumn}
\usepackage{multirow}
\usepackage{makecell}

\begin{document}
\title{Parametric hypersensitivity in many-body bath-mediated transport: The quantum Rabi model}
\author{Chern Chuang}
\affiliation{Chemical Physics Theory Group, Department of Chemistry, and Center for Quantum Information and Quantum
Control, University of Toronto, Toronto, Ontario M5S 3H6, Canada}
\author{Arie Kapulkin}
\affiliation{Department of Physics and Astronomy, Carleton College,
One North College Street, Northfield, MN 55057, USA}
\author{Arjendu K. Pattanayak}
\affiliation{Department of Physics and Astronomy, Carleton College,
One North College Street, Northfield, MN 55057, USA}
\author{Paul Brumer}
\affiliation{Chemical Physics Theory Group, Department of Chemistry, and Center for Quantum Information and Quantum
Control, University of Toronto, Toronto, Ontario M5S 3H6, Canada}
\date{\today}

\begin{abstract}
We demonstrate that non-equilibrium steady states of the dissipative Rabi model show dramatic spikes in transport rates over narrow parameter ranges. Similar results are found for the Holstein and Dicke models. This is found to be due to avoided energy level crossings in the corresponding closed systems, and correlates with spikes in the entanglement entropy of key eigenstates, a signature of strong mixing and resonance among system degrees of freedom. Further, contrasting the Rabi model with the Jaynes-Cummings model reveals this behavior as being related to quantum integrability.
\end{abstract}

\maketitle
\section*{Introduction}
Transport in open quantum systems is of fundamental interest applicable from chemical processes to quantum technologies, where the transported quantity could be energy, charge, or information\cite{GruebeleWolynes2004AccChemRes,PhysRevA.103.062214}. When complicated many-body quantum dynamics are subject to dissipative effects (measurement and environmental fluctuations) and/or external driving, these act to interfere and dephase the wave-like motion, to supply energy for barrier crossing, and to dissipate energy along specific pathways. The results are novel dynamical phenomena not present in closed systems but possible in nonequilibrium steady states (NESS) and in metastable versions of NESS (long-lived trapped states).\cite{Hershfield1993PRL,Anders2008PRL,HsiangHu2015AnnPhys,PhysRevResearch.2.043052}. 

Such NESS are significant and arise, for example, in photophysical and photochemical processes sustained by sunlight, an incoherent light source with turn-on/off time scales much longer than typical relaxation times relevant to energy harvesting and vision\cite{CaoSilbey2009JPCA,JiangBrumer1991JCP,Hoki2011,Mancal2010NJP,OlsinaCao,Tscherbul14,Manzano2016,dodin2019,YangCao2020JPCL,Dodin2021}. Our recent simulations of the steady-state photoisomerization reaction quantum yield of model rhodopsin (first step in vision) show\cite{RetinalChaos1,RetinalChaos2} extremely sensitive dependence on system parameters while the transient dynamics remain nearly unaffected. This is unexpected and important given the ongoing enthusiasm for analyzing biological systems using pulsed laser and monitoring the resulting transient dynamics. Short-time studies miss this hypersensitivity of NESS to system parameters.

Here we expose the roots of such extreme sensitivity with a study of a variety of light-matter interaction models and the corresponding many-body quantum dissipative dynamics. We start with the open quantum Rabi model, which is a two-level system (`spin') bilinearly coupled to a simple harmonic oscillator (`boson'), each sub-system subject to a separate environmental interaction. We find that even this basic model manifests NESS with transport properties that are extremely sensitive to parameter variation. Such behavior also obtains for a number of generalizations detailed in Appendix A. Some unrelated systems\cite{PhysRevLett.109.240402,PhysRevResearch.2.023178} have also presented an interesting, somewhat different transport rate sensitivity, discussed in the Appendix A.  We demonstrate that in each of our cases, a dramatically altered NESS transport in the open system corresponds to low-energy avoided crossings (LAC) in the closed system spectrum, and that the entanglement entropy of key eigenstates spikes at precisely the same parameter values. We also show that this  parametric dependence of NESS is related to the integrability of the system. The Jaynes-Cummings (JC) model, which has greater symmetry than the Rabi model, and is therefore more integrable, does not have this sensitive parametric dependence. Although the main investigation used the Lindblad open quantum system formalism, valid assuming a weakly interacting Markovian environment, similar results were found under generalizations in the nonsecular and multi-bath Redfield formalism framework as detailed in Appendix E.\footnote{In Appendix E it is shown that while hypersensitivity is found in various system-environment interactions and not just in the Lindblad formalism that is used throughout the main presentation, in non-Lindblad scenarios, the parametric dependence of transport rate spikes will generally be different from that of spikes in Lindblad case.}.  

The Lindblad master equation dynamics for the reduced -- i.e. after tracing out environmental degrees of freedom -- density matrix $\rho$ is
\begin{eqnarray}
  \dot{\rho}(t)  &=&\mathcal{L}\rho(t)=(\mathcal{L}_\mathrm{sys}+\mathcal{L}_\mathrm{s}+\mathcal{L}_\mathrm{b})\rho(t) \nonumber\\
       &=&-i\left[H_\mathrm{sys},\rho(t)\right]+\sum_{y=\mathrm{s,b}}L_y\rho(t)L_y^\dagger-\frac{1}{2}\left\{L_y^\dagger L_y,\rho(t)\right\}
    \label{eqn:FullLindblad}
\end{eqnarray}
where $\mathcal{L}$ are the Liouvillian superoperators. The quantum Rabi Hamiltonian is \begin{eqnarray}
    H_\mathrm{sys}&=&\frac{\Delta}{2}\sigma_z+\omega a^\dagger a+\lambda \sigma_x(a^\dagger+a)
\end{eqnarray}
with $\sigma_i$ the Pauli matrices and $a^\dagger$, $a$ the creation and annihilation operators. The parameters $\Delta$ and $\omega$ set energy scales for the spin and the boson degrees of freedom (DOF), respectively, and $\lambda$ is the coupling strength between the two. We are interested in the steady state density operator $\rho_\mathrm{ss}$ for this system under generalized measurement including environmental decoherence. This dissipative coupling is induced by two Lindblad operators $L_\mathrm{s}=\sqrt{r_\mathrm{s}}\sigma_-$ and $L_\mathrm{b}=\sqrt{r_\mathrm{b}}a$, and introduces finite lifetimes ($r_\mathrm{s}^{-1}$ and $r_\mathrm{b}^{-1}$) for excitations of the two DOF. (In the $\lambda\le \min(r_\mathrm{b},r_\mathrm{s})$ limit care must be taken to the use of Eq.~(\ref{eqn:FullLindblad}) as the bath-induced dissipation becomes nonperturbative.) 

We solve Eq.~(\ref{eqn:FullLindblad}) for the steady states $\rho_\mathrm{ss}$ satisfying $\dot{\rho}(t)=0$. This is done by expressing the quantum Rabi Hamiltonian in the number basis of the harmonic oscillator and truncating at $N_b=20$ (Further increasing $N_\mathrm{b}$ yields the same results).  The steady state solution can be found by normalizing the (single) null vector of the corresponding Liouvillian which is also confirmed by numerically propagating a selection of physical initial states to a time scale much longer than $r_\mathrm{s}^{-1}$ and $r_\mathrm{b}^{-1}$. It should be noted that the system ground state is typically not the NESS state, since the local Lindblad formalism does not conform with detailed balance. \footnote{We note that care must be taken in the interpretation of these Lindblad operators as weakly coupling the system to zero temperature spin/boson baths. A thermodynamically consistent treatment of the latter leads to the so-call Redfield equation of motion, which guarantees detailed balance. See further discussion in Appendices D and E.} Transport rates are studied via the steady state spin flux $J_\mathrm{s}\equiv\mathrm{Tr}_\mathrm{s}[\mathrm{Tr}_\mathrm{b}[\mathcal{L}_\mathrm{s}\rho_\mathrm{ss}]\sigma_z]/2$.\footnote{The steady-state spin-down population $P_D = \mathrm{Tr}[|\downarrow\rangle\langle\downarrow|\rho_\mathrm{ss}]$ may also be used to study transport rates, in place of the steady state spin flux $J_\mathrm{s}$.}

\begin{figure}[htbp]
\centering
    \includegraphics[width=14cm]{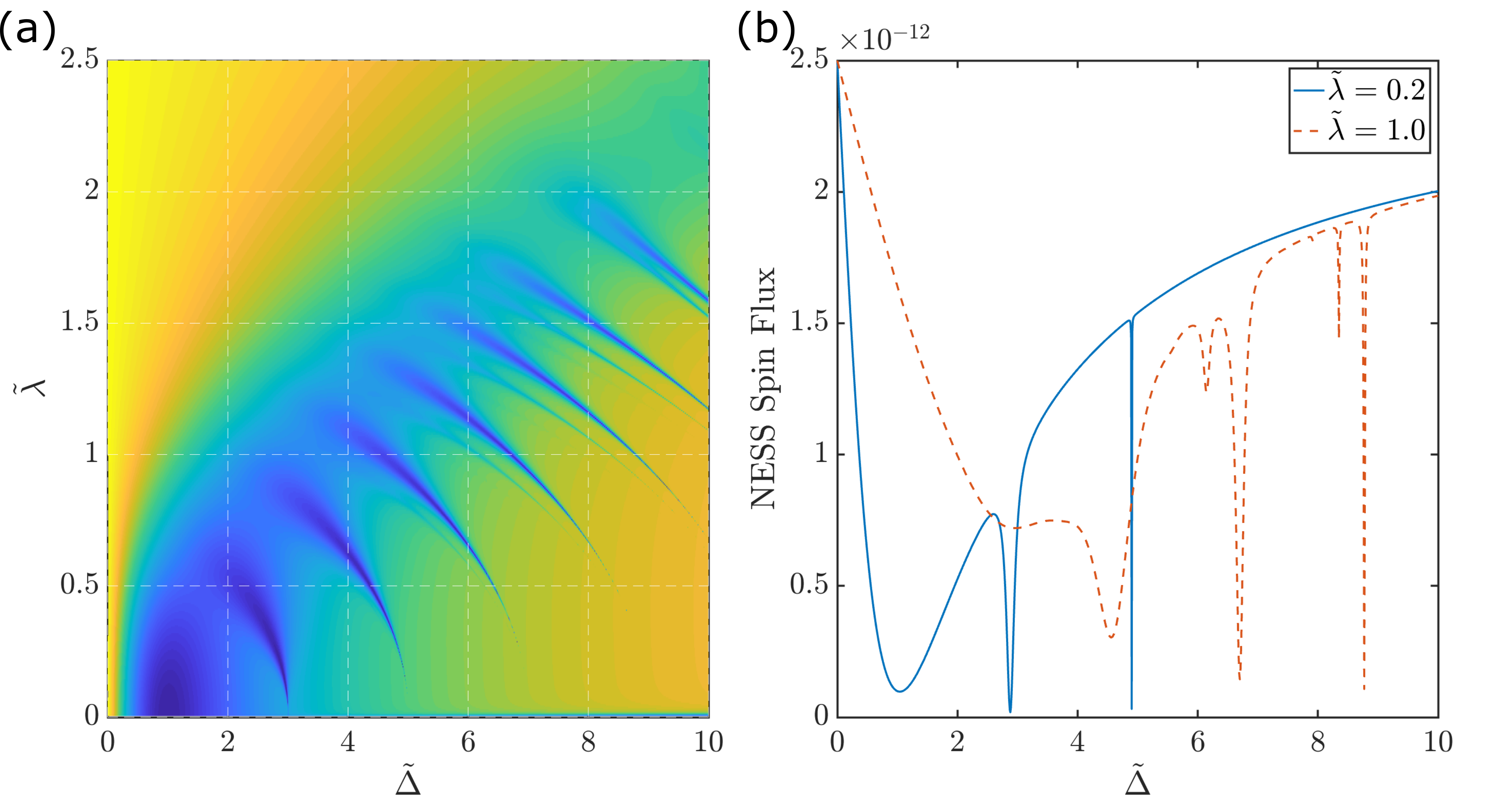}
    \caption{(a) Steady state spin flux $J_\mathrm{s}$ for the open quantum Rabi model as a function of the model parameters $\Tilde{\Delta}=\Delta/\omega$ and $\Tilde{\lambda}=\lambda/\omega$. (b) $J_\mathrm{s}$ at $\Tilde{\lambda}=0.2$ and $\Tilde{\lambda}=1$. The Lindblad rates are set to $(r_\mathrm{s},r_\mathrm{b})/\omega=(5\cdot10^{-12},5\cdot10^{-6})$.
    }
    \label{fig:RabiSS2D}
\end{figure}

Fig.~\ref{fig:RabiSS2D}(a) shows $J_\mathrm{s}$ as a function of the two independent parameters $\Tilde{\Delta}=\Delta/\omega$ and $\Tilde{\lambda}=\lambda/\omega$, where each unique pair $\{\tilde{\Delta},\tilde{\lambda}\}$ corresponds to a different NESS. 
The spin flux is seen to depend in very structured ways on system parameters: (a) there are narrow dips in the limit $\Tilde{\lambda}\rightarrow0$ at odd integer values of $\Tilde{\Delta}$, except for $\Tilde{\Delta}=1$ which shows a broad dip, and this is most pronounced for $\Tilde{\Delta}=3, 5$; (b) as $\Tilde{\lambda}$ increases, dips appear at $\Tilde{\Delta}\approx7, 9, \cdots$ and all dips shift to smaller values of $\Tilde{\Delta}$; (c) the dips flatten with increasing $\Tilde{\lambda}$; (d) secondary and tertiary dips with $\Tilde{\Delta}\ge7$, of lower intensity and narrower line shapes also exist. Two different slices across this figure are plotted in Fig.~\ref{fig:RabiSS2D}(b) showing $J_\mathrm{s}$ as a function of $\Tilde{\Delta}$ for $\Tilde{\lambda}=0.2$, $\Tilde{\lambda}=1$.

We have observed similar sharp parametric sensitivity of transport rates  
in several other many-body open quantum model systems including a generalization of the Rabi model to a few-level system coupled to a harmonic oscillator, the few-atom Dicke model, and the few-site Holstein models, where we used a variety of open system setups. These are summarized in Appendix A. Notably, in the JC model, its few-level and few-spin generalizations, there is no parametric sensitivity. 

\begin{figure}
    \centering
    \includegraphics[width=14cm]{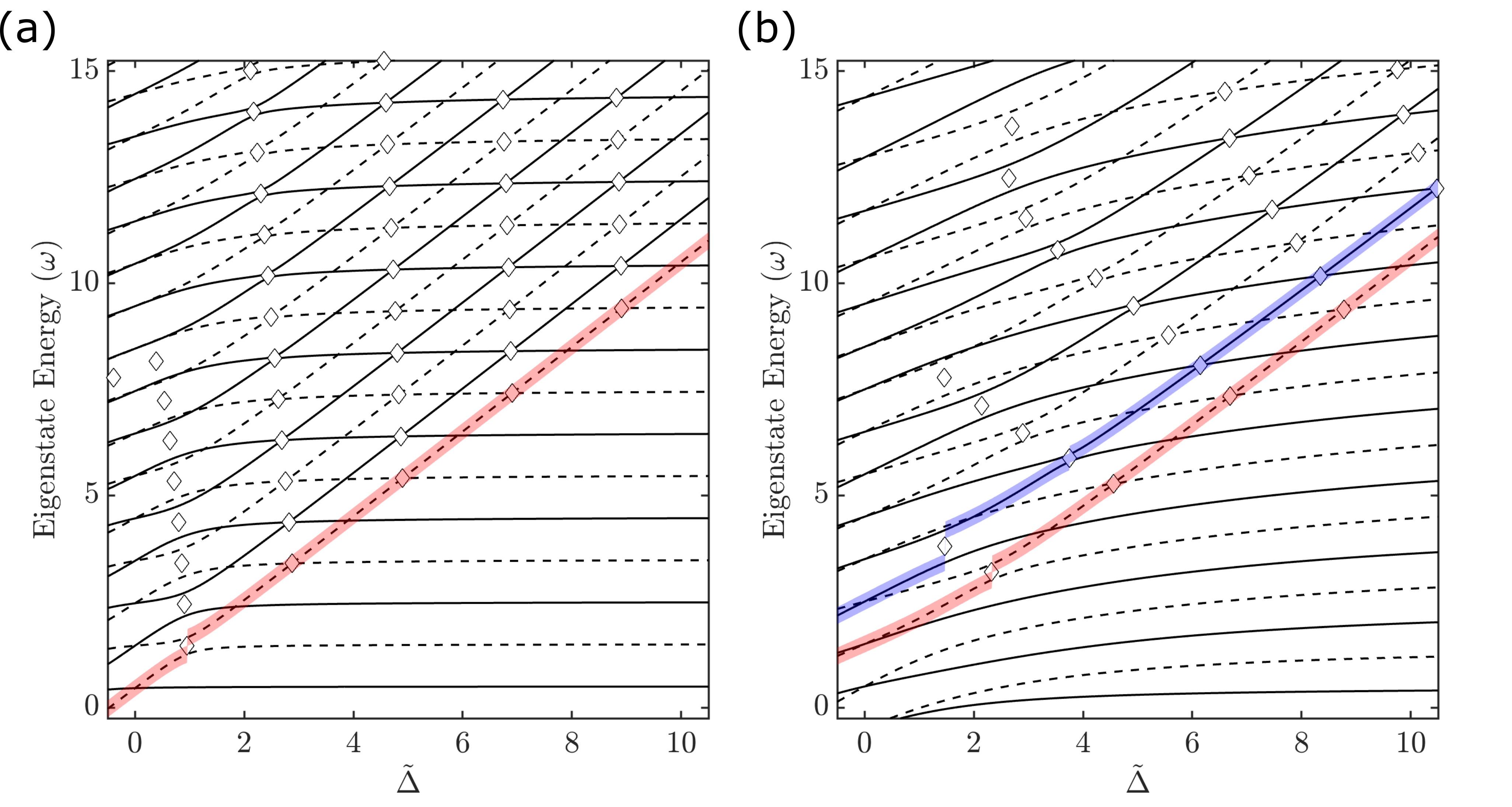}
    \caption{Low-lying energy levels of the Rabi model as functions of $\Tilde{\Delta}$. Solid lines represent positive parity and dashed lines, negative parity. Diamonds mark the locations of energy avoided crossings. (a) $\Tilde{\lambda}=0.2$. (b) $\Tilde{\lambda}=1$.}
    \label{fig:LevelDistribution}
\end{figure}

To unravel the mechanism behind these unusual transport rate dips  we turn to the energy eigenvalues of the corresponding closed many-body system. In Fig.~\ref{fig:LevelDistribution} we show the energy eigenvalues as functions of $\Tilde{\Delta}$ for the same two $\Tilde{\lambda}\in\{0.2,1\}$ slices. This spectrum changes in a complicated fashion with $\Tilde{\Delta}$. It is easier to understand this if we label the eigenvalues according to their parity (dashed lines are positive parity and solid lines are negative parity), which is the eigenvalue of the parity operator $\hat{P}=\sigma_z\otimes\sum_n(-1)^n|n\rangle\langle n|$, where $|n\rangle=\frac{\left(a^\dagger\right)^n}{\sqrt{n!}}|0\rangle$ is the $n$th excited state of the boson DOF. The focus on parity is motivated by a critical difference between the JC and Rabi models which we discuss in more detail below. For the Rabi model, we see directly the important role of parity: lines of different parities tracing eigenvalues as functions of system parameters can cross freely (are degenerate at crossings) while those of the same parity avoid crossings (are prevented from degeneracy); these avoided crossings are indicated by diamonds. 

The parametric dips in transport rates in Fig.~\ref{fig:RabiSS2D} occur at the LAC in Fig.~\ref{fig:LevelDistribution}. The LAC for $\Tilde{\lambda}=0.2$ are at odd integer values of $\Tilde{\Delta}$, as are the dips of $J_\mathrm{s}(\tilde{\Delta})$. Further, the secondary structure of $J_\mathrm{s}(\tilde{\Delta},\tilde{\lambda})$ is also related, since the secondary dips for $J_\mathrm{s}(\tilde{\Delta},\Tilde{\lambda}=1)$ in Fig.~\ref{fig:RabiSS2D}(b) coincide with the \textit{second lowest} energy avoided crossings in Fig.~\ref{fig:LevelDistribution}(b). Lastly, the prominence of the dips correlates with the size and sharpness of the avoided crossings, as defined by either dip width (e.g. comparing the widths of the three dips at $\Tilde{\Delta}\approx1, 3, 5$ for $\Tilde{\lambda}=0.2$) or dip magnitude (e.g. comparing the main and the secondary dips for $\Tilde{\lambda}=1$). 
Thus, the parametric spikes are seen to arise due to resonant coupling between the two DOF comprising the same parity eigenstates of the closed system, which are then exploited by the open system dynamics.

Only the LAC contribute, which is consistent with our understanding that for a zero-temperature environment only the low-lying part of the spectrum matters. Notably, as shown in Fig.~\ref{fig:LevelDistribution}(a) \textit{which} specific pair of states \textit{are} the LAC is dependent on $\Tilde{\Delta}$. 

\begin{figure}
    \centering
    \includegraphics[width=14cm]{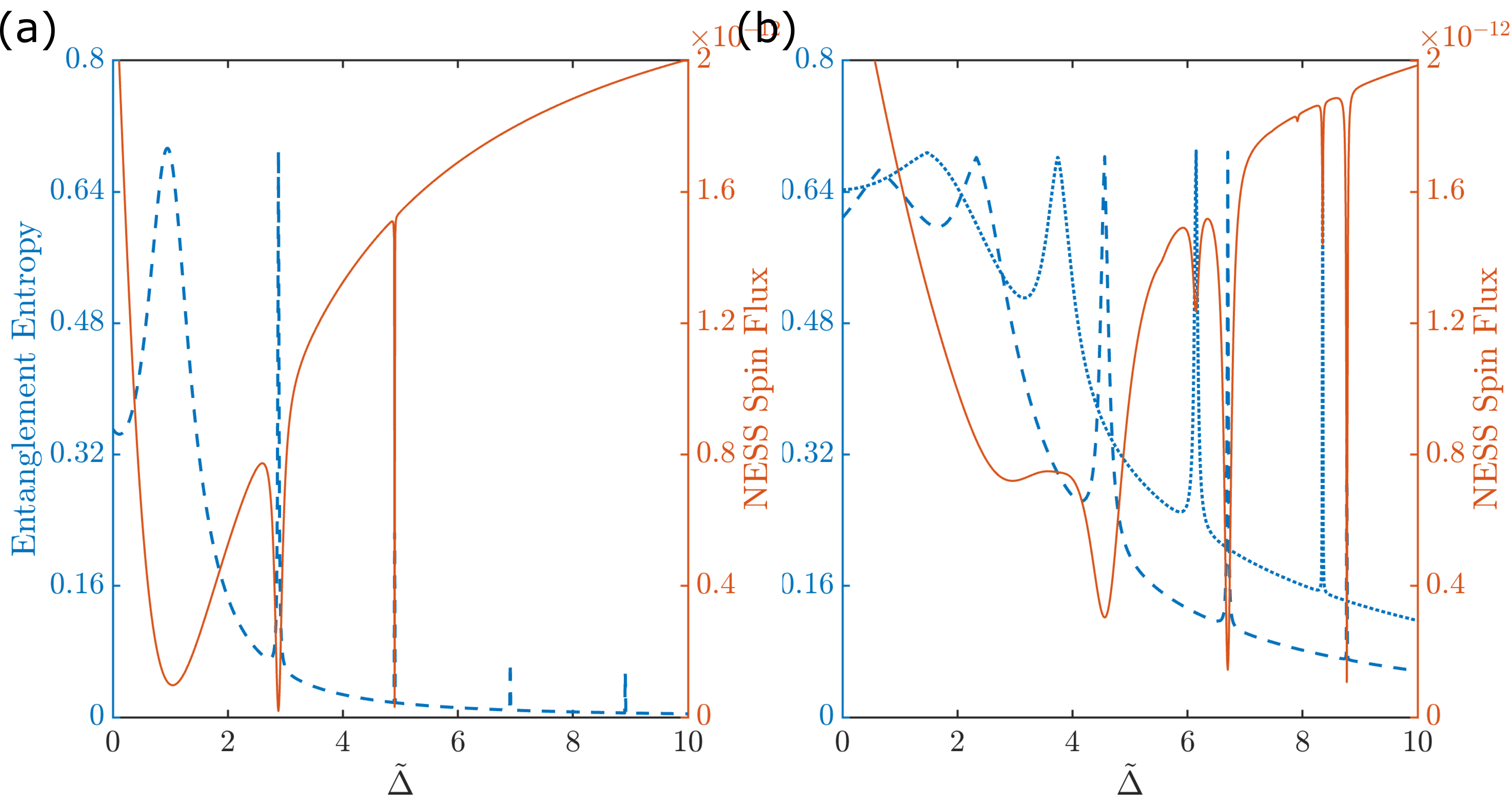}
    \caption{The entanglement entropy of sequences of energy eigenstates as functions of $\Tilde{\Delta}$ (dashed and dotted lines).  The corresponding steady state spin flux $J_\mathrm{s}(\tilde{\Delta})$ is also shown (solid lines). (a) $\Tilde{\lambda}=0.2$, the states correspond to those highlighted in red in Fig.~\ref{fig:LevelDistribution}(a). (b) $\Tilde{\lambda}=1$, the dahsed and dotted lines correspond to the states highlighted in red and blue respectively in Fig.~\ref{fig:LevelDistribution}(b).}
    \label{fig:Rabi_EntanglementEntropy}
\end{figure}

To further probe the quantitative connection between open system transport and closed system properties we consider the entanglement entropy of the LAC eigenstates of the Rabi model. The entanglement entropy measures the mixing in a given state among various DOF of the system. For a pure state $|k\rangle$ in the Hilbert space of the quantum Rabi model it is natural to define the entanglement entropy through
\begin{eqnarray}
S_k=-\mathrm{Tr}_\mathrm{b}[\rho_{k,\mathrm{s}}\log\rho_{k,\mathrm{s}}]=-\mathrm{Tr}_\mathrm{s}[\rho_{k,\mathrm{b}}\log\rho_{k,\mathrm{b}}]
\end{eqnarray}
where $\rho_{k,\mathrm{s}}=\mathrm{Tr}_\mathrm{b}|k\rangle\langle k|$ and $\rho_{k,\mathrm{b}}=\mathrm{Tr}_\mathrm{s}|k\rangle\langle k|$. By construction, a product state has zero entanglement entropy whereas finite entropy indicates mixing, with its magnitude indicating the strength of the resonant interaction between the spin and the boson DOF.
Fig.~\ref{fig:Rabi_EntanglementEntropy}(a) shows $S_k$ for the red LAC state in Fig.~\ref{fig:LevelDistribution}(a), compared to the NESS spin flux $J_\mathrm{s}(\tilde{\Delta})$. We find excellent agreement between the dip locations for the two. Fig.~ \ref{fig:Rabi_EntanglementEntropy}(b) highlights both these primary dip states as well as the sequence of states tracking the \textit{second} lowest energy avoided crossings (all of negative parity), indicated by the blue line in Fig.~\ref{fig:LevelDistribution}(b), associated with the secondary dips shown in Fig.~\ref{fig:RabiSS2D}.
For $\Tilde{\lambda}=1$ in particular the primary dips coincide with the spikes in entanglement entropy of the lowest LAC state (red in Fig.~\ref{fig:LevelDistribution}(b)) while the secondary dips coincide with that of the second LAC state (blue).  It is also worth noting that in the large $\Tilde{\lambda}$ regime where the NESS flux dips flatten (e.g.  in the range of $0<\Tilde{\Delta}<4$ for $\Tilde{\lambda}=1$ in Fig.~\ref{fig:Rabi_EntanglementEntropy}(b)) there exist overlapping resonances with broad lineshapes in terms of $S_k$.  This evidences that potential transport sensitivity can be masked by an overabundance of overlapping resonances within a phase space region. 

Thus, the delocalization of LAC many-body states correlates with observed transport sensitivity in the open system. While this single $S_k$ measure of the LAC states provides information about the transport `resonance', it is not trivial to generalize, especially as the number of DOF increase.  There, for example, a subset of system DOFs might come into resonance (resulting in one or multiple avoided crossings) that are immaterial to the transport in other subspaces of the Hilbert space, particularly given the critical role of the chosen Lindblads. Here, an appropriate bipartition for the entanglement entropy could be designed to match with the transport in question: For example if transport takes place primarily within a single DOF then a coarse-grained measure can be taken within the said DOF\cite{RahavJarzynski2007JStatMech} to assess the degree of wavefunction delocalization across the reactant-product barrier. Studies of these issues, as well as the fact that NESS transport anomaly due to level avoided crossings also manifests as deviation from an appropriate thermal state in an quantum thermodynamics perspective\cite{Poletti2019Entropy} are in progress.

Our previous work of a model photochemical reaction found parametric sensitivity in the presence of Wignerian nearest-neighbor energy level spacing distribution, an established measure of quantum chaos in closed systems.\cite{RetinalChaos1} At the same time, a normal-to-superradiant quantum phase transition has been demonstrated to occur in the quantum Rabi model in the $\Tilde{\Delta}\rightarrow\infty$ limit at the critical coupling strength $\lambda_c=\sqrt{\Delta\omega}/2$.\cite{Plenio2015PRL} This transition has been recently associated with the onset of quantum chaos as measured by out-of-time-order correlation functions.\cite{PhysRevA.105.032444} In the present work, the Rabi model is well within the normal phase, and we conclude that a quantum chaotic closed system -- in either sense above -- is \textit{not} necessary for parametric hypersensitivity in open system transport rates.  Instead, each individual spike/dip in the transport rate is associated with a resonant level avoided crossing which not necessarily belongs to a Wignerian-distributed system. 

\begin{figure}
    \centering
    \includegraphics[width=14cm]{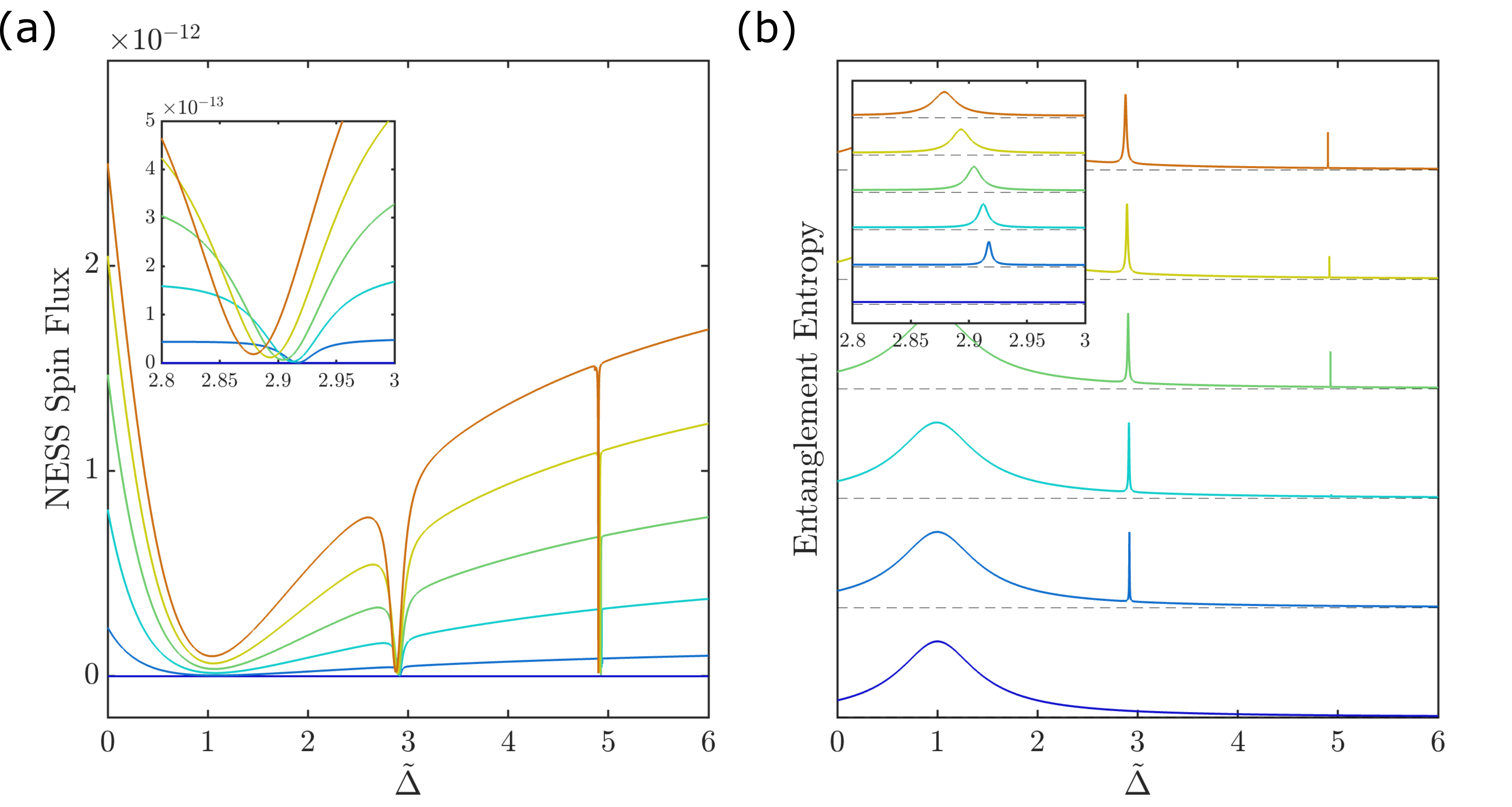}
    \caption{(a) $\Tilde{\Delta}$-dependence of $J_\mathrm{s}$ for various values of $s$ in Eq.~(\ref{eqn:JC2Rabi}). $\Tilde{\lambda}$ is set to 0.04. Red is the Rabi model ($s=1$) and blue the JC model ($s=0$), with intermediate colors in increments of $\Delta s=$0.2. (b) Entanglement entropy of the sequence of energy eigenstates defined similar to those highlighted in Fig.~\ref{fig:LevelDistribution}(a). Successive $s$ values are shifted vertically by unity for clarity. The insets zoom in on the vicinity of $\Tilde{\Delta}=3$.}
    \label{fig:JC2Rabi}
\end{figure}

Consider, however, the role of integrability and conserved quantities (including parity).\cite{PhysRevE.77.066202,Braak2011PRL} The Rabi model, whose Hamiltonian commutes with the Parity operator, is solvable.\cite{Braak2011PRL} Its rotating-wave approximated version, the JC model, has more operators that commute with the Hamiltonian, which has been exploited to yield closed form expressions for eigenvalues.\cite{PhysRevA.91.053808} Since the open version of the quantum Rabi model shows parametric hypersensitivity while that of the JC model does not, we examine the role of integrability via a system with tunable counter-rotating terms
\begin{eqnarray}
    H_\mathrm{sys}(s)&=&\frac{\Delta}{2}\sigma_z+\omega a^\dagger a+\lambda [\sigma_-a^\dagger+\sigma_+a+s(\sigma_-a+\sigma_+a^\dagger)]
    \label{eqn:JC2Rabi}
\end{eqnarray}
where $\sigma_\pm=\sigma_x\pm i\sigma_y$ and $0\le s\le 1$\cite{PhysRevE.77.066202}, yielding the JC model at $s=0$ and the Rabi model at $s=1$. It should be noted that for the JC model all the avoided crossings become level crossings except for those at $\Tilde{\Delta}=1$, representing the level repulsion between the so-called JC doublets. In Fig.~\ref{fig:JC2Rabi} (a) we show $J_\mathrm{s}(\tilde{\Delta})$ as in Fig.~\ref{fig:RabiSS2D}(b), with $s$ linearly scaling from 0 (blue) to 1 (red). No $\Tilde{\Delta}$-dependence is observed for the JC model, and the parametric sensitivity for $s\ne0$ grows as $s$ increases. The rotating wave approximation holds best at $\Tilde{\Delta}\approx1$. Indeed, we find the least difference between the NESS solutions to the two models in this regime. 

This preliminary study therefore indicates that many-body open system transport is affected by the integrability of the system through its effect on avoided crossings. In Fig.~\ref{fig:JC2Rabi}(b) we map $S_k$ for the the relevant LAC energy eigenstates defined as before\footnote{For the JC model ($s=0$)  we pick the sequence of eigenstates following the lowest energy level crossings instead. This sequence of states are the $s\rightarrow0$ limits of $s\ne0$ sequences examined in other cases.} and again see correlation between entanglement entropy and transport. However, for the JC model, there is a broadened $\Tilde{\Delta}=1$ peak for the entanglement entropy while there is vanishing open system flux regardless of $\Tilde{\Delta}$. Thus, avoided crossings are not a sufficient condition for open system rate sensitivity. However, we conjecture that it is a \textit{necessary} condition, i.e. that all parametric spikes in open system transport are correlated with an avoided crossing in the closed system arising from a resonance among its DOF. This has held true across all of our investigations.

In conclusion: We have examined the parametric sensitivity of NESS transport rates in the open quantum Rabi model, a prototypical model for light-matter interaction, and linked its appearance to the avoided crossings of energy eigenvalues of the corresponding closed system arising from resonances between the DOF within the system. For the zero-temperature dissipative environment studied, the dips in the transport rates can be individually mapped to specific low-energy avoided crossings. Several other few-body quantum systems, as well as several variations of dissipative environments show similar behavior. Further, the entanglement entropy of selected eigenstates of the closed many-body Hamiltonian is useful in predicting parametric sensitivity in the open system. We work deep in the so-called `normal regime' of the Rabi model in the context of quantum phase transition so that in this case it is clear that quantum chaos is not intimately related to parametric hypersensitivity.  Finally, our study of Hamiltonians that interpolate between the JC model and the Rabi model that have different degrees of quantum integrability indicates that some form of quantum integrability is key in determining such parametric sensitivity in open quantum systems.

\section*{Acknowledgement}
This work was supported by the U.S. Air Force Office of Scientific Research (AFOSR) under grant number FA9550-21-1-0098. This work has benefited greatly from discussions with Professor Jennifer Ogilvie (University of Michigan) on parametric hypersensitivity in biophysical systems.

\renewcommand\thefigure{A.\arabic{figure}} 
\setcounter{figure}{0}
\renewcommand{\theequation}{A.\arabic{equation}}
\setcounter{equation}{0}
\section*{Appendex A: Other Quantum Model Systems}
In addition to the Rabi and the Jaynes-Cummings models discussed in the main text,  we also examined a number of different quantum models commonly employed in the study of quantum physics.  They are listed below and tabulated in Table.\ref{tab:models}.  They can be considered as generalizations of the Rabi model,  composed of discrete degree(s) of freedom (DOF) coupled to an harmonic DOF.   In Table.\ref{tab:models} we not only summarize the existence of sensitivity by lowest avoided crossings (LAC) but also their locations in the weak inter-particle coupling limit ($\tilde{\lambda}\rightarrow0$).  For example,  in the Rabi model the LAC-induced sensitivity is located at odd integer $\tilde{\Delta}$ values in this limit. 

In all of the cases, we apply the same Lindblad formalism to these model and solve for the nonequilibrium steady state (NESS), $\partial\rho/\partial t=0$.  Detailed analysis of the correspondence between the LAC and the transport sensitivity will be documented in a forthcoming report.  Here we briefly summarize the results of symmetry broken Rabi and Jaynes-Cummings models, which are the most relevant to our current discussion.

\begin{table}[]
\begin{tabular}{|c|c|c|c|c|}
\hline
Model                                     				& JC		                         		& Rabi					& JC, Broken $U(1)$    			& Rabi, Broken $\mathbb{Z}_2$ 		\\ \hline
Sensitivity                               			& No                          			& Yes                           		& Yes                               			& Yes                       		\\ \hline
Resonance ($\tilde{\lambda}\rightarrow0$) 	& -						& Odd Integer $\tilde{\Delta}$	& $\tilde{\Delta}=0$       			& Integer $\tilde{\Delta}$        	\\ \hline\Xhline{2\arrayrulewidth}\hline
Model                                     				& Multi-level JC			& Multi-level Rabi             		& Dicke Dimer/Trimer           		& Holstein $N$mer ($N=2,3,4$) 		 \\ \hline
Sensitivity                               			& No 					& Yes                          			& Yes                          				& Yes                     		                            \\ \hline
Resonance ($\tilde{\lambda}\rightarrow0$) 	& - 	& Odd Integer $\tilde{\Delta}$\footnote{The main peaks, there are more subtle ones.} 						& Odd Integer $\tilde{\Delta}$ 		& $\tilde{\Delta}=\frac{n}{m}$\footnote{$n,m=0, 1,\cdots,N-1$}                            \\ \hline
\end{tabular}
\label{tab:models}
\end{table}

\subsection*{Symmetry Broken Rabi and Jaynes-Cummings Models}
\begin{figure}
    \centering
    \includegraphics[width=7cm]{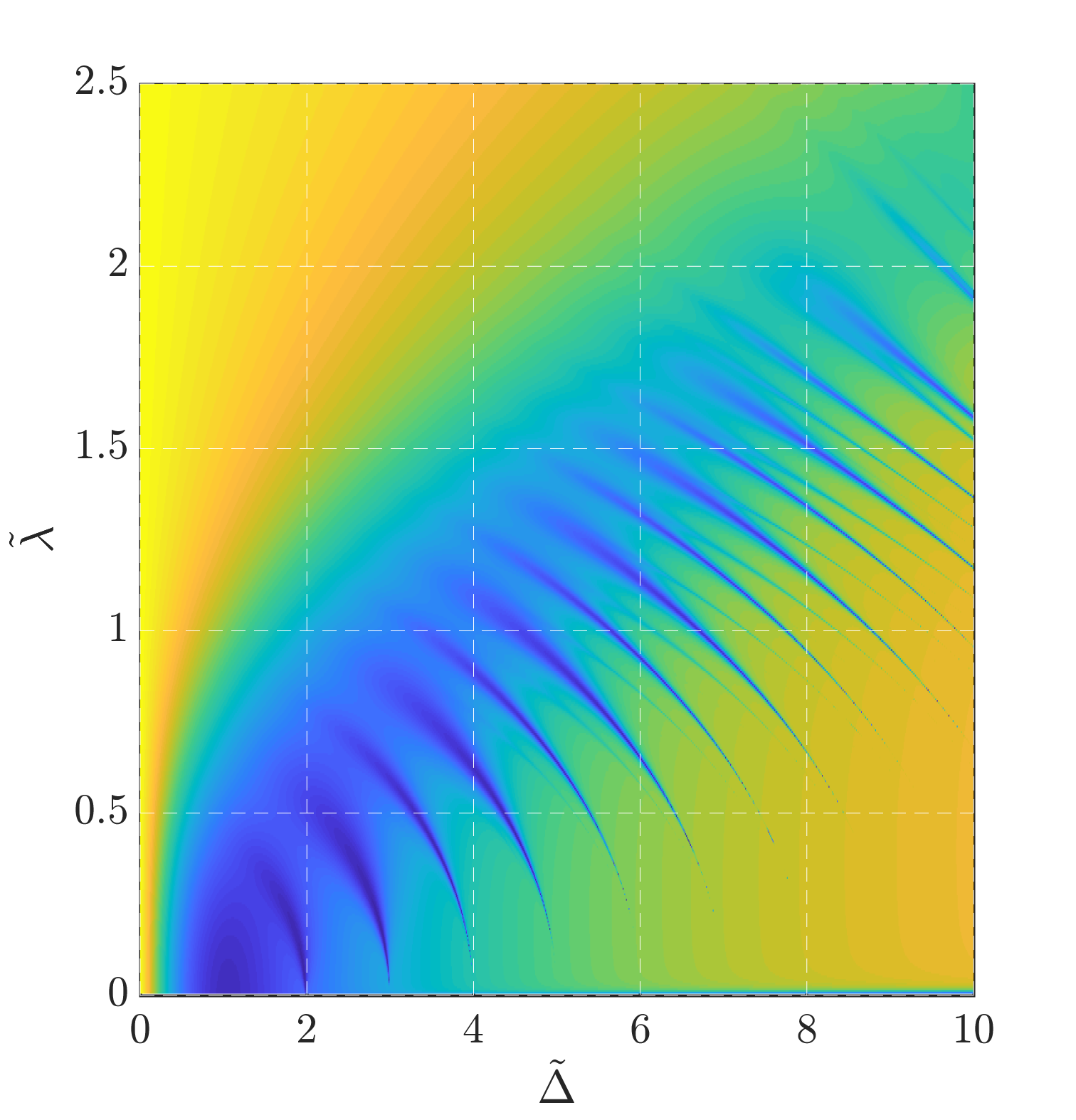}
    \includegraphics[width=7cm]{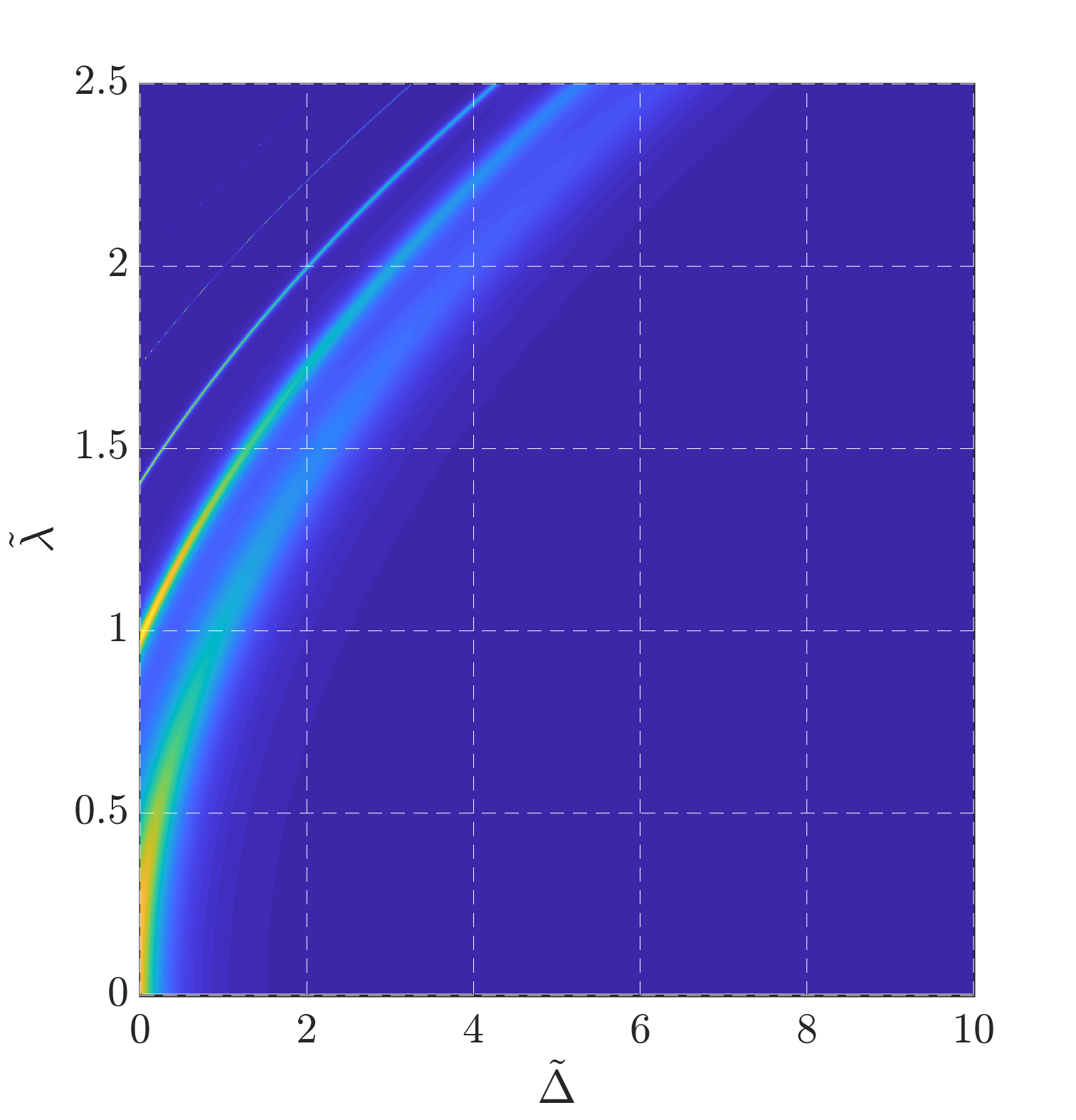}
    \caption{Spin flux at NESS,  taking $\epsilon/\omega=0.15$,  $r_\mathrm{s}/\omega=5\cdot10^{-12}$, and $r_\mathrm{b}/\omega=5\cdot10^{-6}$.  (Left) Broken-$\mathbb{Z}_2$ Rabi model. (Right) Broken-$U(1)$ Jaynes-Cummings model. }
    \label{fig:BrokenZ2}
\end{figure}

In Braak's examination of the quantum integrability of the Rabi model,  he proposed the addition of a $\mathbb{Z}_2$ symmetry breaking term $\epsilon\sigma_x$ that renders the model nonintegrable. \cite{Braak2011PRL} We apply the same modification to the base Rabi and Jaynes-Cummings models while keeping the same open system setup and parameters (Lindbladian) and calculate the NESS.  The results are shown in Fig. ~\ref{fig:BrokenZ2}.

In the broken symmetry Rabi model,  the resonance occurs at all integer values of $\tilde{\Delta}$ in the weak spin-boson coupling limit, whereas in the original Rabi model it is restricted to odd integers of $\tilde{\Delta}$. This is a direct consequence of the symmetry breaking which makes the LAC between unlike-parity states at the even $\tilde{\Delta}$ values.  In the case of broken symmetry Jaynes-Cummings model the effect is even more dramatic,  where resonance sensitivity comes into existence from where there is none.  Note that in this case the larger symmetry group of the Jaynes-Cummings model,  $U(1)$,  is broken.  This renders the model nonintegrable in Braak's sense since the only quantum number is the energy. \cite{Braak2011PRL} 

\subsection*{NESS Transport in Many-Spin Systems in the Literature} Several previous studies of many-spin systems reported parametric dependence of NESS transport. For example, in Ref.\cite{PhysRevLett.109.240402} parametric dependence of NESS transport was shown to signify quantum phase transitions at critical points. This can be considered as a special case as the LAC responsible for the parametric dependence is that between the ground and the first excited states. It is worth noting that additional information on the system character under this scenario can be extracted if a thermodynamically consistent treatment of the environment is carried out.\cite{PhysRevResearch.2.023178} 

On the other hand, from a quantum thermal engine point of view, it has been shown that NESS transport through spin systems with nontrivial geometry shows parametric dependence that correlates to the magnitude of the deviation from thermal states.\cite{Poletti2019Entropy} In other word, this suggests that the maximum amount of extractable work in this situation, referred to as ergotropy, is also closely related to the behavior of NESS transport and its parametric dependence. 

\renewcommand{\theequation}{B.\arabic{equation}}
\setcounter{equation}{0}

\section*{Appendix B: System-Bath Parameter Dependence}
In the main text and up to this section in the Appendices we have been focusing on the variation of system parameters while keeping the same system-bath parameters, specifically the dissipation rates $r_\mathrm{s}$ and $r_\mathrm{b}$.  While a more detailed exploration and analysis of the form of the dissipator,  the level of treatment for open quantum systems, as well as their effects on the resonance sensitivity are left for future study,  we briefly examine the effect of changing the relative dissipation rate $(r_\mathrm{s},r_\mathrm{b})$. 

\begin{figure}
    \centering
    \includegraphics[width=7.0cm]{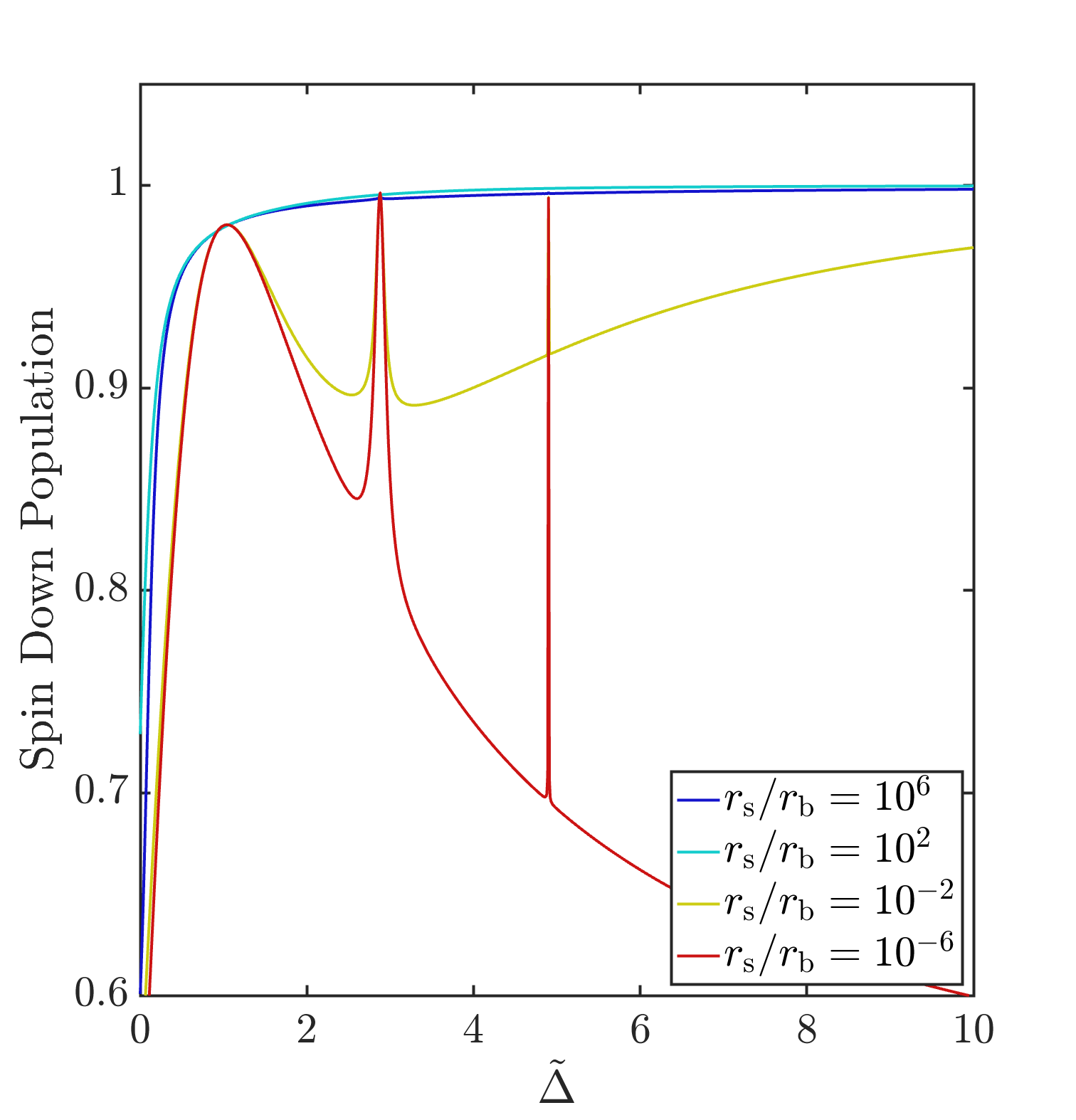}
    \includegraphics[width=7.0cm]{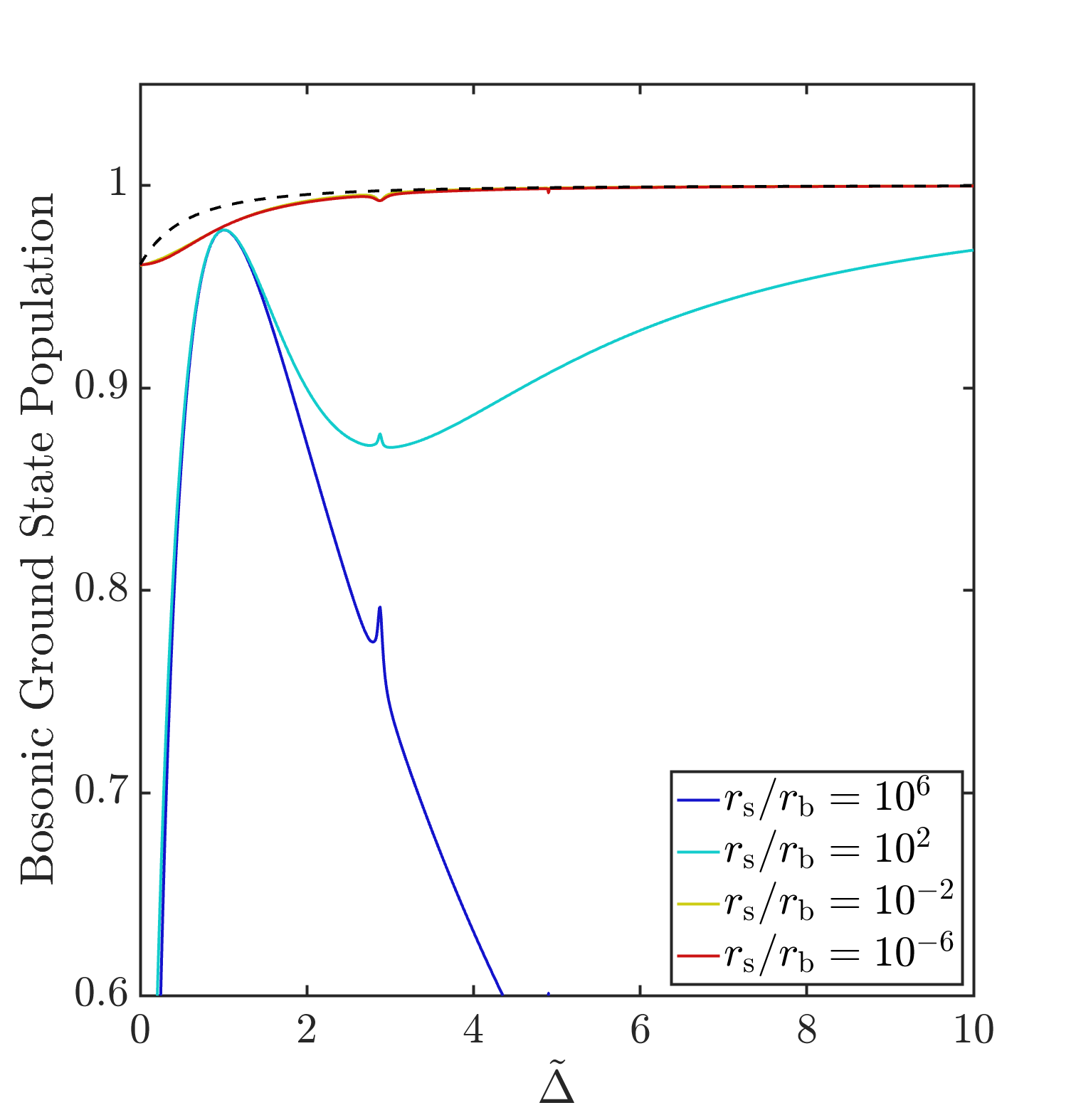}
    \caption{(Left) Spin down population as a function of $(r_\mathrm{s},r_\mathrm{b})$ and $\tilde{\Delta}$, $\tilde{\lambda}=0.2$ for Rabi model.  From red to blue: $(r_\mathrm{s},r_\mathrm{b})=5\cdot(10^{-12},10^{-6}),~5\cdot(10^{-11},10^{-7}),\cdots5\cdot(10^{-6},10^{-12})$. (Right) The lowest bosonic level population. }
    \label{fig:r1r2Dep}
\end{figure}

The results are shown in Fig.~\ref{fig:r1r2Dep}. Here all system parameters are kept the same except for $\tilde{\Delta}$, whose value is scanned for comparing the resonance sensitivity between different sets of dissipation rates $(r_\mathrm{s},r_\mathrm{b})$.  The LAC-induced peaks in the NESS are seen in all cases, both in terms of the spin down population as well as the boson ground state population. Their locations are the same (odd integer $\tilde{\Delta}$) regardless of the values of the dissipation rates. However,  it is clear that the prominence of peaks is sensitive to $(r_\mathrm{s},r_\mathrm{b})$. 

Specifically,  the peak prominence for spin population is most significant when $r_\mathrm{s}/r_\mathrm{b}\ll1$. The opposite is true for that of the boson population.  This can be rationalized by considering that with or without the LAC-induced sensitivity, intra-system transport takes place owing to the interplay between the system Hamiltonian and the dissipation. However, the sensitivity is most dramatic when the transport is the slowest among the competing kinetic processes. Taking the NESS spin down population as example,  LAC sensitivity is minimal when $r_\mathrm{s}$ is large, since it is not the bottleneck for the kinetic network described by the Lindblad equation of motion. Consequently, regardless of the existence of LAC the transport of spin population is unhindered. Conversely if $r_\mathrm{s}$ is small, then LAC can make a large impact on the transport.  We expect this general principle to apply to other systems and more general setup of NESS transport.

\renewcommand{\theequation}{C.\arabic{equation}}
\setcounter{equation}{0}
\section*{Appendix C: Secular Approximation to Lindbladian}
In simulating the dynamics as well as the steady state solution to the Lindblad equation of motion, we construct the corresponding Liouvillian superoperators of the size $N^2$-by-$N^2$, where $N$ is the dimension of the system. With larger systems this quickly becomes unmanageable. However, in the case of extreme weak system-environment coupling, \textit{i.e.}, $|\sum_y L^\dagger_yL_y|\ll|H_\mathrm{sys}|$ 
the time scale of pure system evolution is much faster than that of the transitions induced by the Lindblad terms. In this limit, one is justified the secular approximation which decouples the dynamics of the population and coherence in the system energy eigenbasis. (Note that this is different from the usual secular approximation applied to Redfield dynamics, which turns it into a Lindblad form. )  This significantly reduces the numerical effort and results in a master equation for the eigenstate population that takes a simple form as follows.
\begin{eqnarray}
\frac{\partial P^\mathrm{(e)}}{\partial t}&=&\sum_yW_yP^\mathrm{(e)}\\
\langle m^\mathrm{(e)}|W_y|n^\mathrm{(e)}\rangle&=&|L_{y,mn}^\mathrm{(e)}|^2-\delta_{mn}\sum_k|L_{y,kn}^\mathrm{(e)}|^2
\end{eqnarray}
where the superscript (e) indicates system energy eigenbasis.  The coherence terms $\rho_{mn}^\mathrm{(e)}$ oscillate at the corresponding system energy gap $E_m-E_n$ and decay at the rate of $\sum_y(W_{y,mm}+W_{y,nn})/2$. 

\begin{figure}
    \centering
    \includegraphics[width=7.0cm]{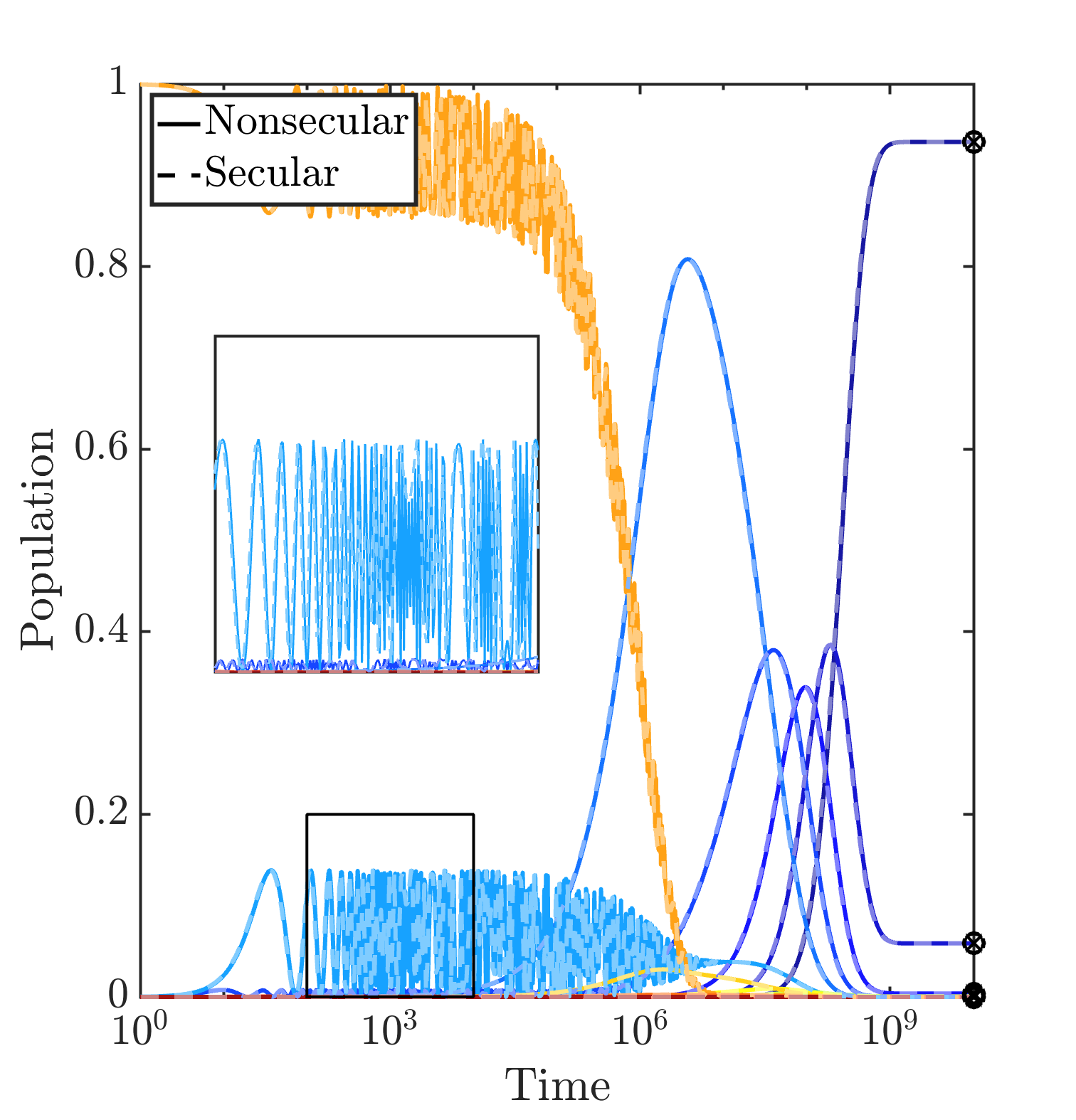}
    \includegraphics[width=7.0cm]{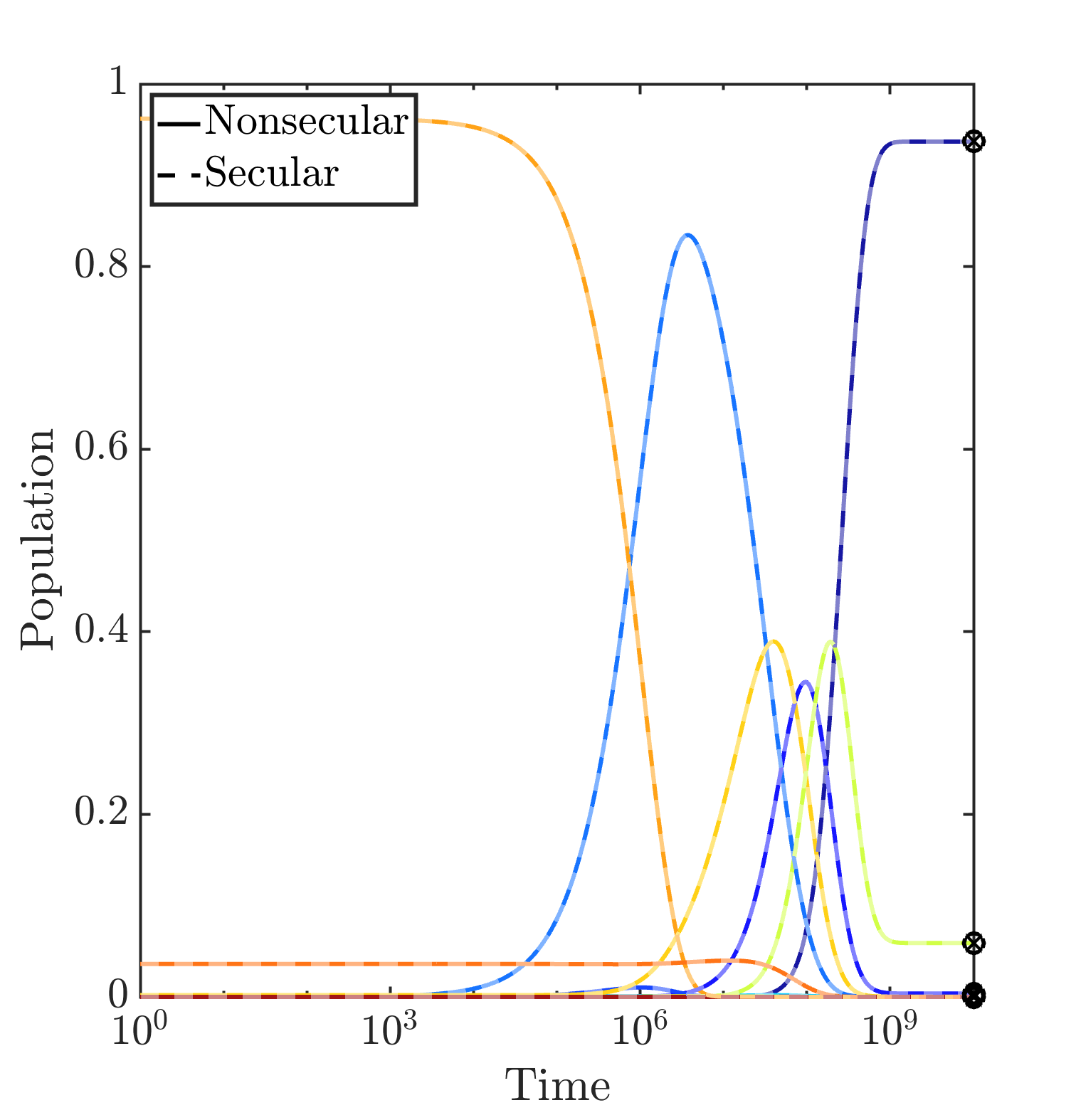}
    \caption{The temporal dynamics of a typical Rabi model system with $(\tilde{\Delta},\tilde{\lambda})=(0.6,0.05)$, truncated at $N_\mathrm{b}=10$. The Lindblad rates are set to $(r_\mathrm{s},r_\mathrm{b})/\omega=5\cdot(10^{-12},10^{-6})$.  The initial state is a product state of spin up and boson Fock state $|n_\mathrm{b}=5\rangle$.  The NESS solutions are shown as symbols to the right of each panel: nonsecular (circles) and secular (crosses). (Left) Population dynamics in the site (direct product state) basis. (Right) Population dynamics in the energy eigenbasis.}
    \label{fig:secular}
\end{figure}

In Fig.~\ref{fig:secular} we compare the dynamics of full nonsecular Lindblad equation of motion and its secular approximated counterpart in both the site (spin-boson product state, left) basis and in the energy eigenbasis (right). First, notice that in all cases the temporal dynamics converges to the corresponding NESS solution (symbols). Second, the NESS solutions are insensitive to the secular approximation.  While minute deviation between the two trajectories can be detected in the transient regime and in the site basis (inset),  they are in general near identical in all other regimes. On the other hand, for observables that depend on the eigenstate coherence this method becomes unreliable and full nonsecular Lindblad calculation is needed.

\renewcommand{\theequation}{D.\arabic{equation}}
\setcounter{equation}{0}
\section*{Appendix D: Steady State Population and Deviation from Thermal State}
In the main text we present the spin flux as the metric for parametric sensitivity. Similarly, once the NESS has been obtained given the system parameters, we find that the NESS population also reveals the same trend. In particular, since we are adopting an effective zero temperature Lindblad equation of motion, it is noteworthy that the corresponding NESS population shows parametric sensitivity. This is because thermal states (in our case the ground state), $\rho(T)=e^{-H_\mathrm{sys}/T}/Z$ where $Z$ is the partition function,  do not exhibit sensitivity toward parameter variation in all of our simulations. While the fundamental reasons for the discrepancy between the two states are explored and discussed in the next section, here we compare the deviation of NESS from the ground state by numerically solving the Lindblad equation of motion.

\begin{figure}
    \centering
    \includegraphics[width=7.0cm]{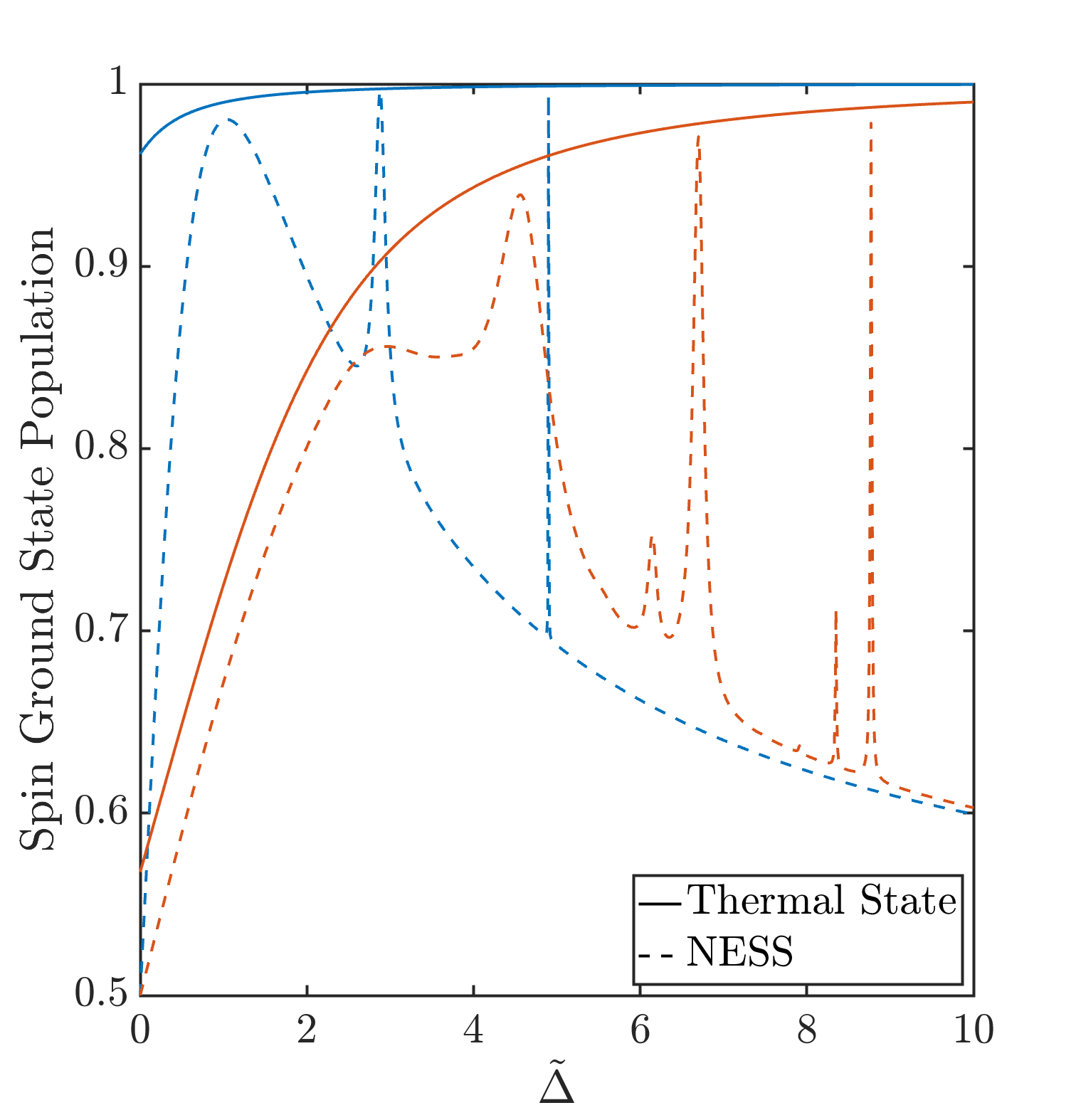}
    \includegraphics[width=7.0cm]{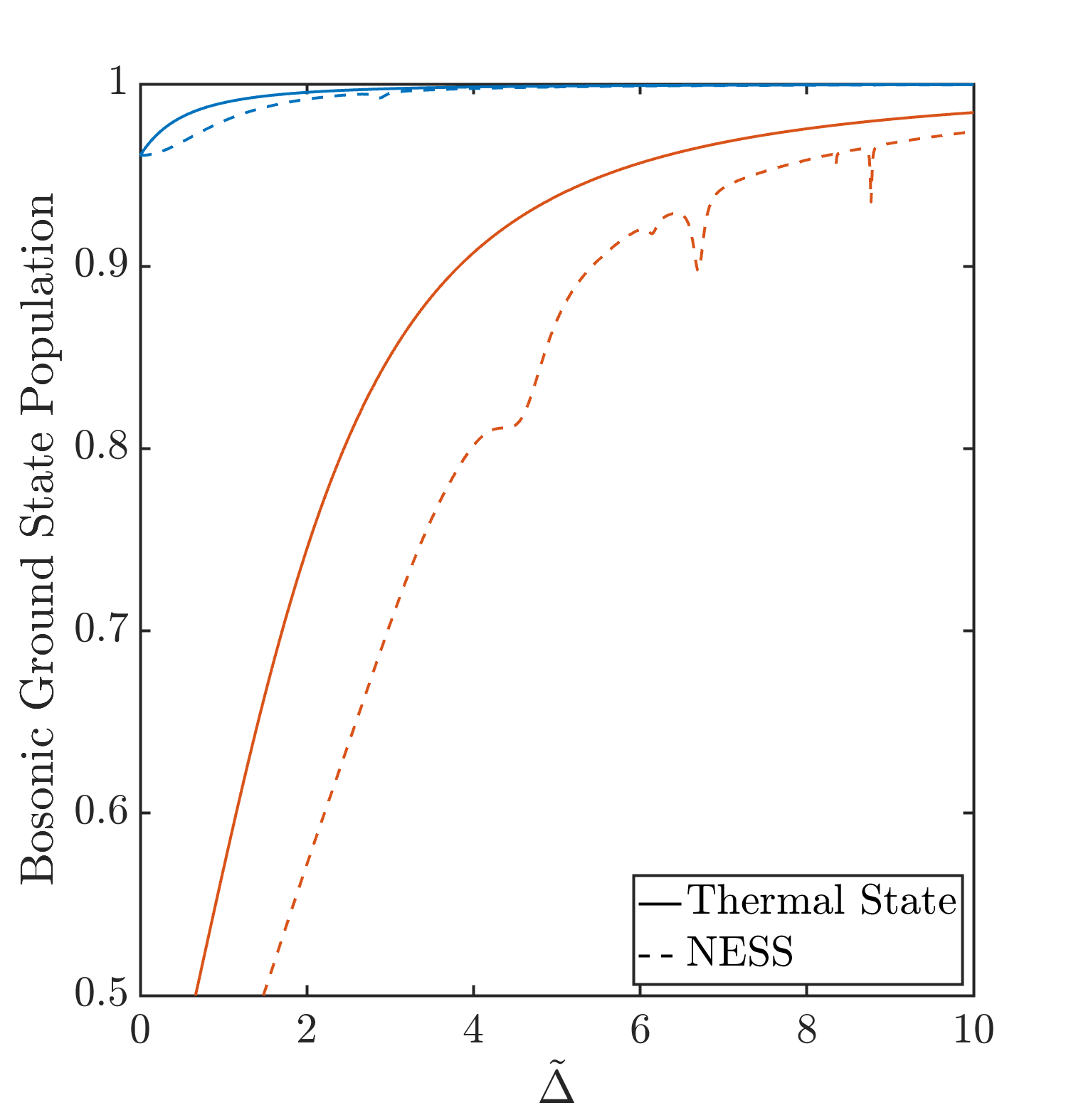}
    \caption{(Left) Spin ground state populations as functions of $\tilde{\Delta}$ for $\tilde{\lambda}=0.2$ (blue) and $\tilde{\lambda}=1$ (brown). The Lindblad rates are $(r_\mathrm{s},r_\mathrm{b})/\omega=5\cdot(10^{-12},10^{-6})$. The thermal states are shown in solid lines and the NESS states are in dashed lines. (Right) Bosonic ground state populations. }
    \label{fig:CmpThermalState}
\end{figure}

In Fig.~\ref{fig:CmpThermalState} we show both the NESS ($\rho_\mathrm{ss}$) and the ground state ($\rho(T=0)$) populations of the ``local'' states as functions of $\tilde{\Delta}$: the spin ground state ($|\downarrow\rangle$) and the bosonic ground state ($|N=0\rangle$).  While the two states deviate throughout the parameter range covered, it is clear that the difference is minimized at the locations of LAC, where the resonance effectively brings the NESS close to the corresponding thermal state. This is expected as LAC mark the maximal mixing of spin and bosonic DOF so that the effect of local dissipation/driving (as in our case) is most delocalized. We will return to this point in the next section.

\renewcommand{\theequation}{E.\arabic{equation}}
\setcounter{equation}{0}
\section*{Appendix E: Redfield Equation of Motion: Detailed Balance and Thermodynamic Consistency}
It has been argued that at finite inter-particle coupling strengths ($\tilde{\lambda}$),  the effect of environmental induced relaxation cannot be properly handled by the Lindblad dynamics discussed so far.  \cite{PhysRevA.84.043832,Agarwal_2013,Plenio2018PRA} In fact with finite $\tilde{\lambda}$, the scattering rates among the system energy eigenstates do not obey the detailed balance relation. Consequently, the system does not relax to the thermal state $e^{-\beta H_\mathrm{sys}}/Z$ when the multiple baths are kept at the same temperature $T=\beta^{-1}$.

While it has been shown that a Lindblad-type dissipator like the ones adopted in this study can be realized effectively by driving a system in certain ways (specifically in a driven trapped ion system\cite{Plenio2018PRA} or in cavity-assisted Raman scattering experiments\cite{PhysRevA.87.033814}), it is of interest to examine how the LAC-induced transport sensitivity takes place under a thermodynamically consistent treatment of the system in contact with passive baths without driving.  To this end, we adopt a generic open quantum system model, where the system is the Rabi model coupled to two sets of quantum harmonic oscillators via its spin and boson DOF, separately and independently. 
\begin{eqnarray}
H&=&H_\mathrm{sys}+H_\mathrm{bath}+H_\mathrm{sb}=H_\mathrm{sys}+\sum_y[\omega_{y,k}a^\dagger_{y,k}a_{y,k}+g_{y,k}(a^\dagger_{y,k}+a_{y,k})S_y],
\end{eqnarray}
where $S_y$ is a system operator. Specifically we take $S_\mathrm{s}=\sigma_x$ and $S_\mathrm{b}=a^\dagger_{k}+a_{k}$.  The summation over $k$ for the bath modes is, in fact, an integration over continuous, bounded analytic functions $J_y(\omega_y)$ typically referred to as the bath spectral densities. The specific form of the spectral density is immaterial in our case, here chosen to be of an Ohmic form $J_y(x)=\eta_yx e^{-x/\gamma_{y}}$. Here $\eta_y$ is the coupling strength and $\gamma_{y}$ is the cutoff frequency, the latter chosen to match the respective system energy scales, \textit{i.e.} $\gamma_\mathrm{s}=\Delta$ and $\gamma_\mathrm{b}=\omega$.  

The effects of these independent baths are then treated separately using the standard Markovian Redfield formulation,\cite{Ishizaki2009} and the final Redfield kernel is the summation of those of the two baths.  The final equation of motion can be written as
\begin{eqnarray}
\frac{\partial\rho_{mn}^\mathrm{(e)}}{\partial t}&=&-i\epsilon_{mn}\rho_{mn}^\mathrm{(e)}+\sum_y\mathcal{R}_{y;mn,op}\rho_{op}^\mathrm{(e)}\\
\mathcal{R}_{y;mn,op}&=&-\delta_{np}\sum_{q}\Gamma_{y,mqqo}+\Gamma_{y,pnmo}+\Gamma_{y,pnmo}^*-\delta_{mo}\sum_q\Gamma_{y,pqqn}^*.
\end{eqnarray}
Here $\epsilon_{mn}$ is the energy gap between eigenstates and 
\begin{eqnarray}
\Gamma_{y,mnop}=\langle m^\mathrm{(e)}|S_y|n^\mathrm{(e)}\rangle\langle o^\mathrm{(e)}|S_y|p^\mathrm{(e)}\rangle J_y(\epsilon_{op})n_\mathrm{BE}(\epsilon_{op},T_y)
\end{eqnarray}
where $n_\mathrm{BE}(x)$ is the Bose-Einstein distribution and $T_y$ is the temperature of the bath. For detailed discussion on the applicability of Lindblad treatment of perturbative baths, see Ref.\cite{PhysRevA.105.032208}.

\begin{figure}
    \centering
    \includegraphics[width=7.0cm]{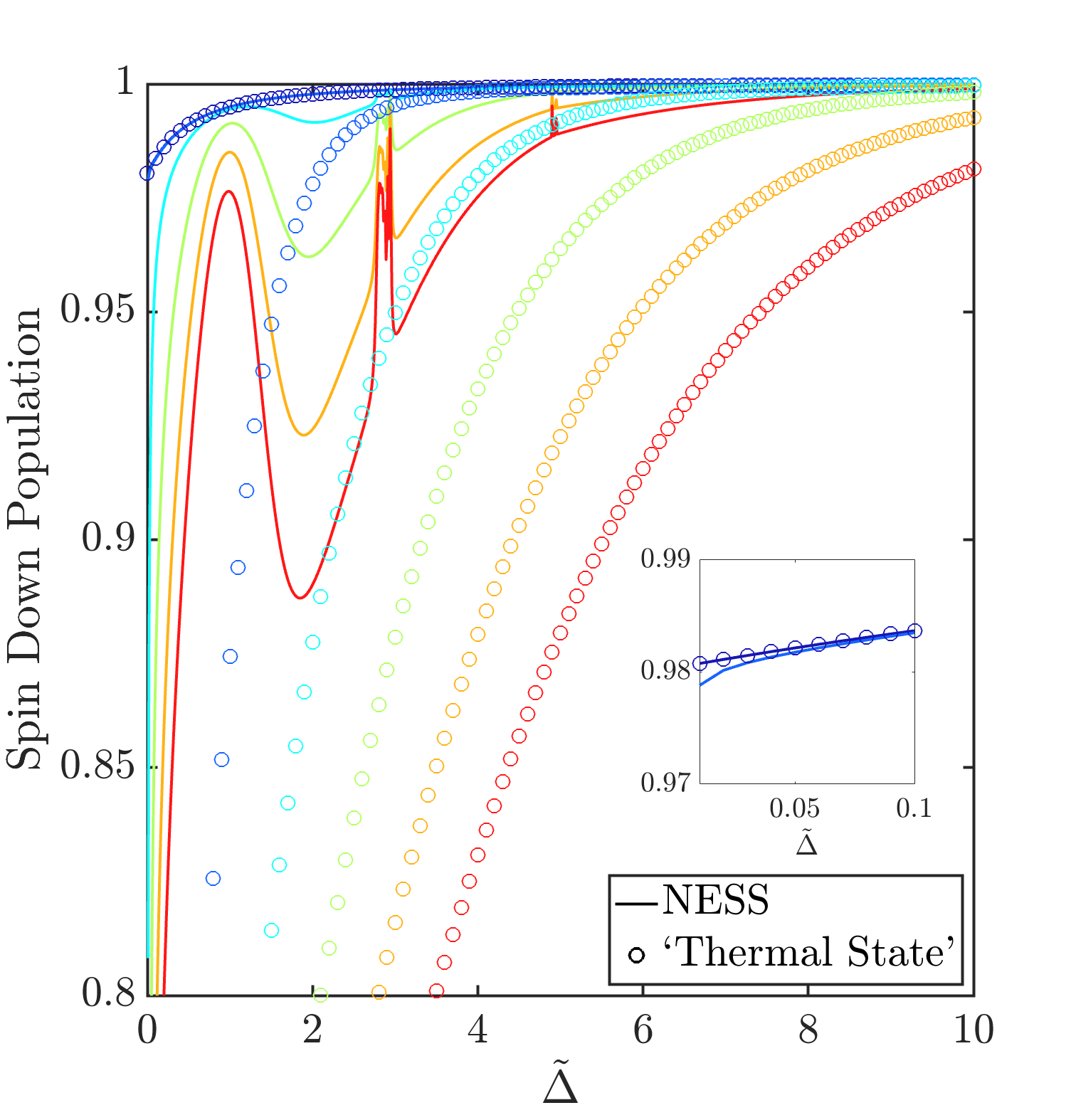}
    \includegraphics[width=7.0cm]{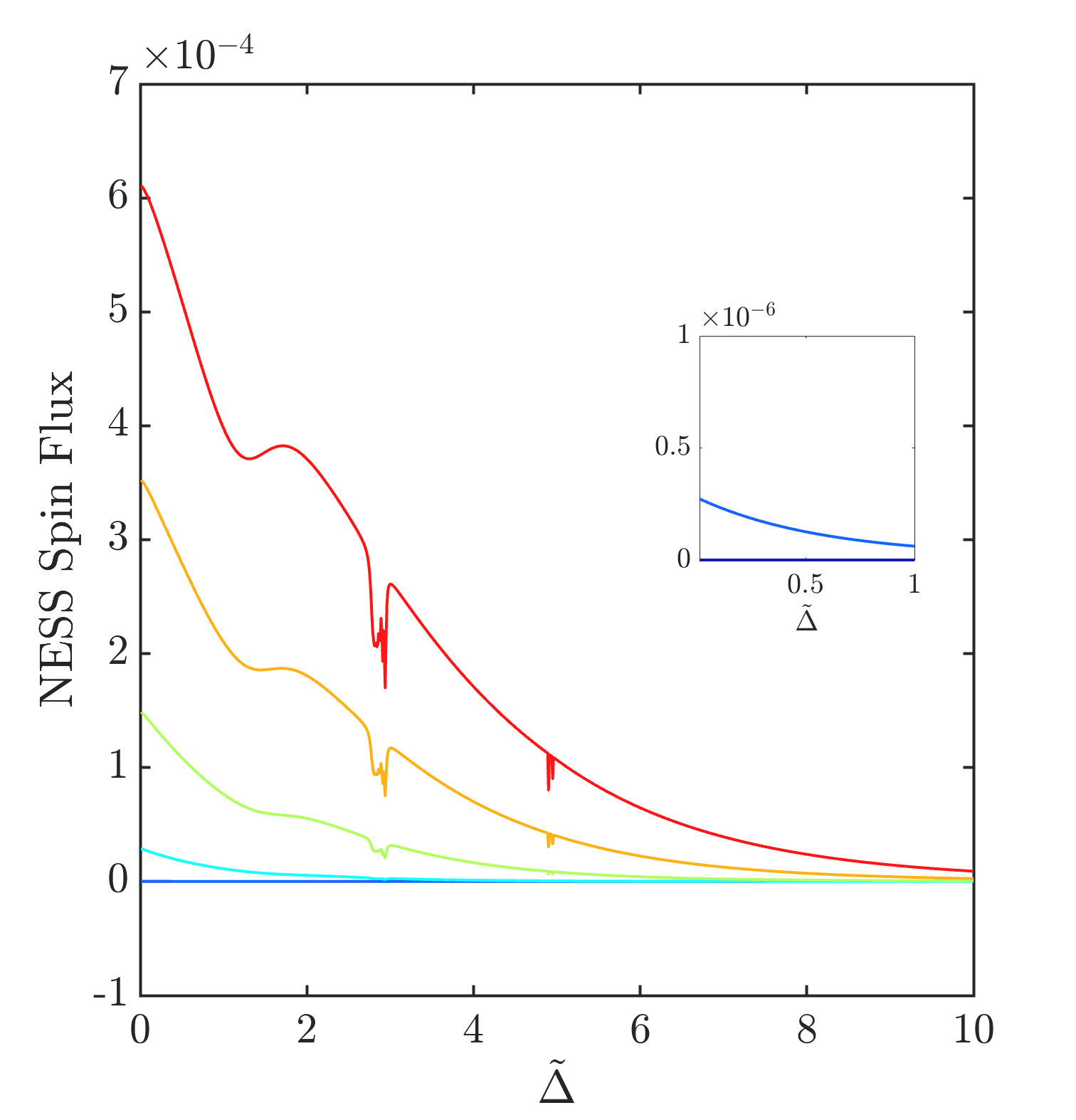}
    \caption{(Left) Spin down NESS population for Rabi model ($\tilde{\lambda}=0.2$) coupled to two different temperature baths treated with secular Redfield method (solid). The dashed lines represent the thermal state population at the spin bath temperature. The spin bath temperature is varied from $T_\mathrm{s}/\omega=0$ (blue) to $T_\mathrm{s}/\omega=2.5$ (red) with linear spacing. (Right) The corresponding spin flux at NESS as a function of spin bath temperature. The insets of both panels zoom in on the small $\Tilde{\Delta}$ regime to better distinguish $T_\mathrm{s}/\omega=0$ and $0.5$ results. }
    \label{fig:RedfieldNESS}
\end{figure}

We first examine the NESS solution to the Redfield equation of motion where we keep the boson bath at zero degrees and vary the spin bath. In Fig.~\ref{fig:RedfieldNESS}, we show the $\tilde{\Delta}$ dependence of both the spin down population and the NESS spin flux as functions of the spin bath temperature. (color-coded)  Also, we deliberately set the system-bath coupling strengths $\eta_y$ such that $\eta_\mathrm{s}/\eta_\mathrm{b}=0.01$, \textit{i.e.} weaker system-spin bath coupling than the boson bath, for reasons mentioned in the discussion of Fig.~\ref{fig:r1r2Dep}.  For the population we also plot the thermal state population in dashed lines $\left[\rho(\beta_\mathrm{s})\propto e^{-\beta_\mathrm{s} H_\mathrm{sys}}\right]$, where $\beta_\mathrm{s}$ is the inverse temperature of the spin bath).

A few points are notable in analysing the results shown in Fig.~\ref{fig:RedfieldNESS}. First, the NESS obtained by solving the additive Redfield baths identifies with the thermal state when the two baths are at the same temperature, i.e. $\rho_\mathrm{ss}\propto \exp[-H_\mathrm{sys}/T]$. This is a direct consequence of the detailed balance requirement for the Redfield tensor under the secular approximation.  In this situation no LAC-induced features are observed, as the thermal state is a smooth function of system parameters regardless of resonances.  Second, with increasing temperature bias the deviation from the thermal states of both baths develops, and the associated steady state flux is proportional to the bias.  Finally, LAC-induced parametric sensitivity becomes more prominent with increasing temperature bias.  This shows that the observed parametric sensitivity is not an artifact due to the simple Lindblad treatment without accounting for the inter-particle coupling terms in the system Hamiltonian.  However, further investigation using non-perturbative treatment of the baths is needed to assess the generality of this phenomenon in the intermediate/strong system-bath coupling regime. 

\begin{figure}
    \centering
    \includegraphics[width=7.0cm]{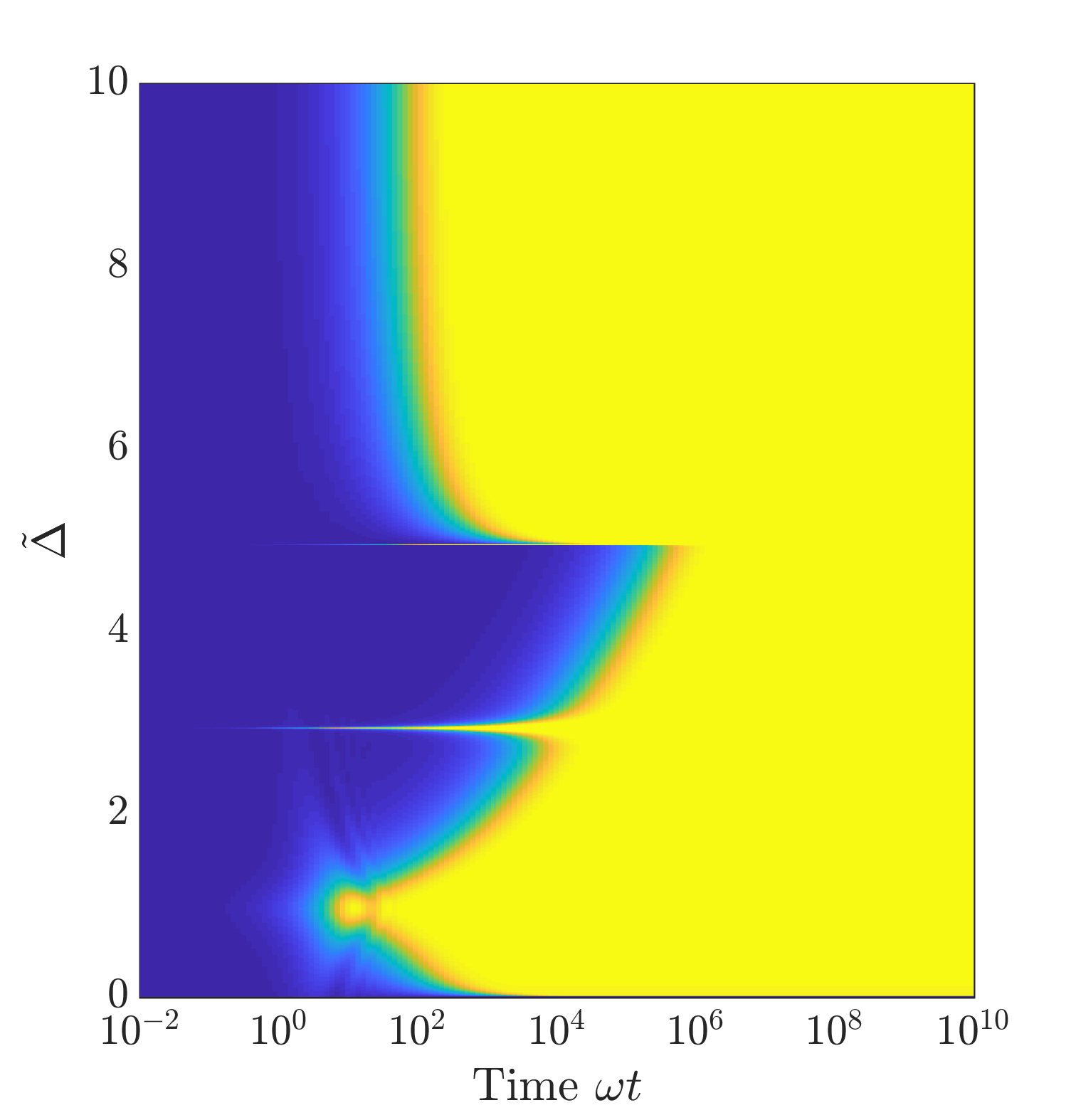}
    \includegraphics[width=7.0cm]{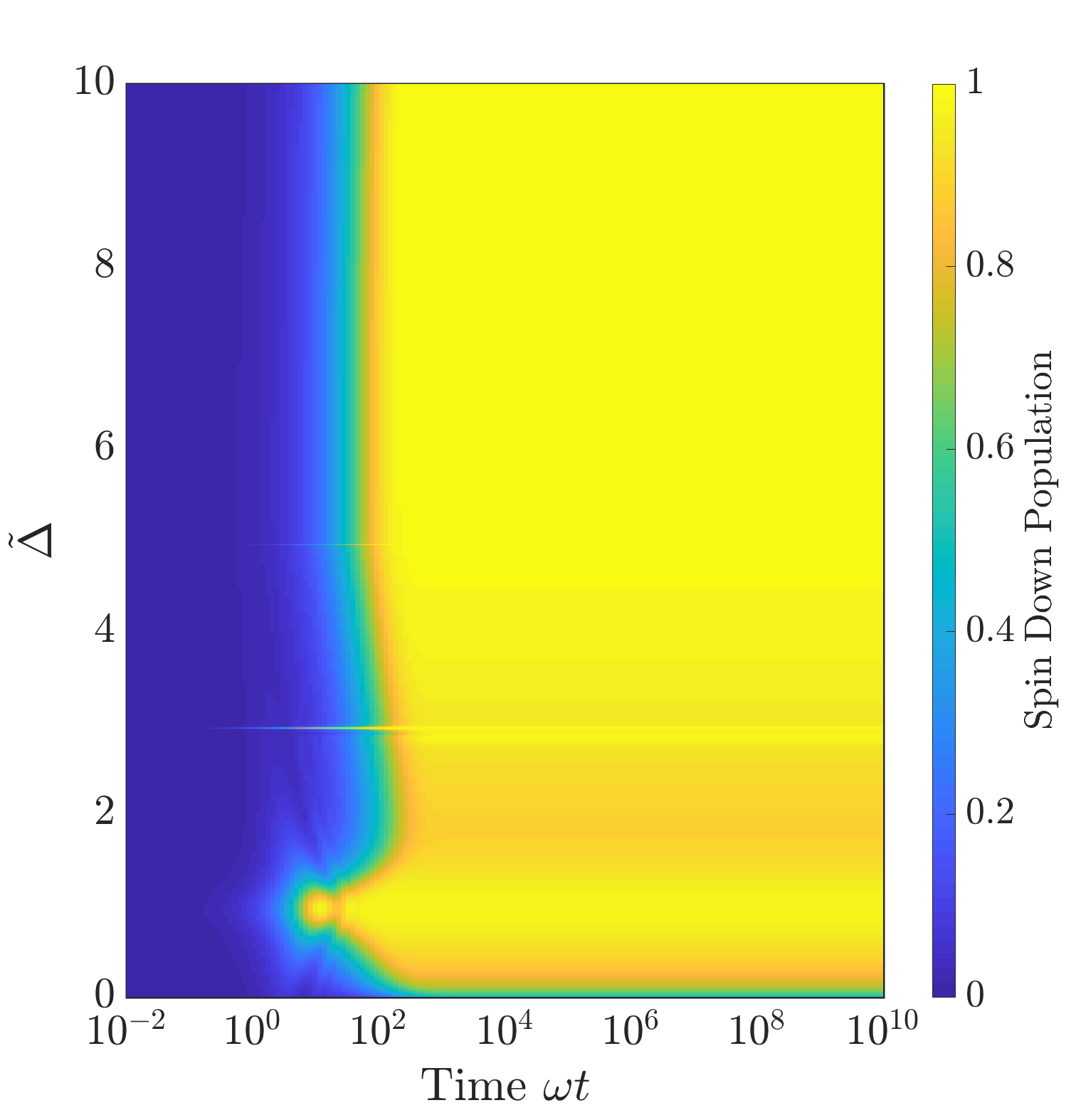}
    \includegraphics[width=7.0cm]{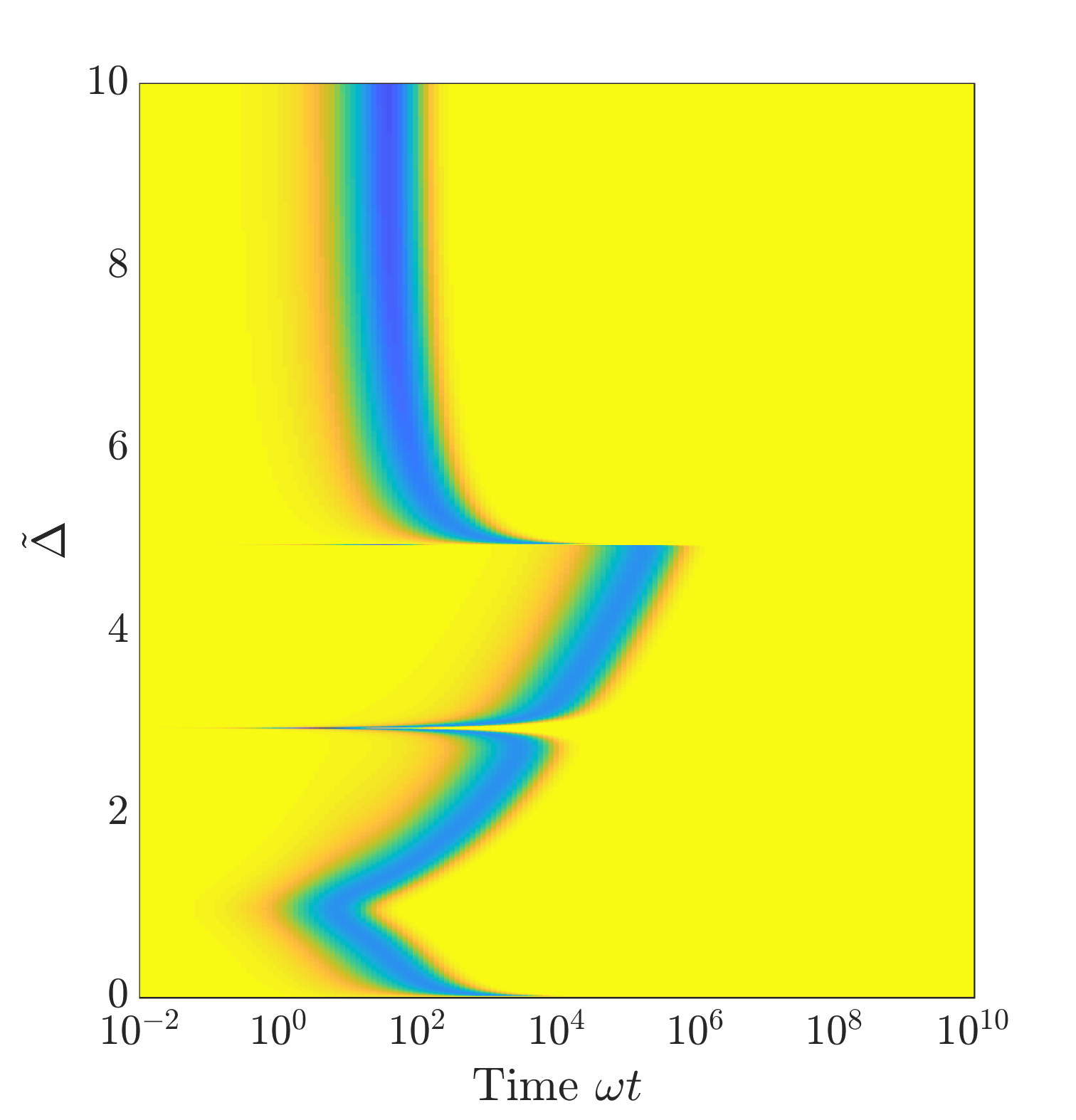}
    \includegraphics[width=7.0cm]{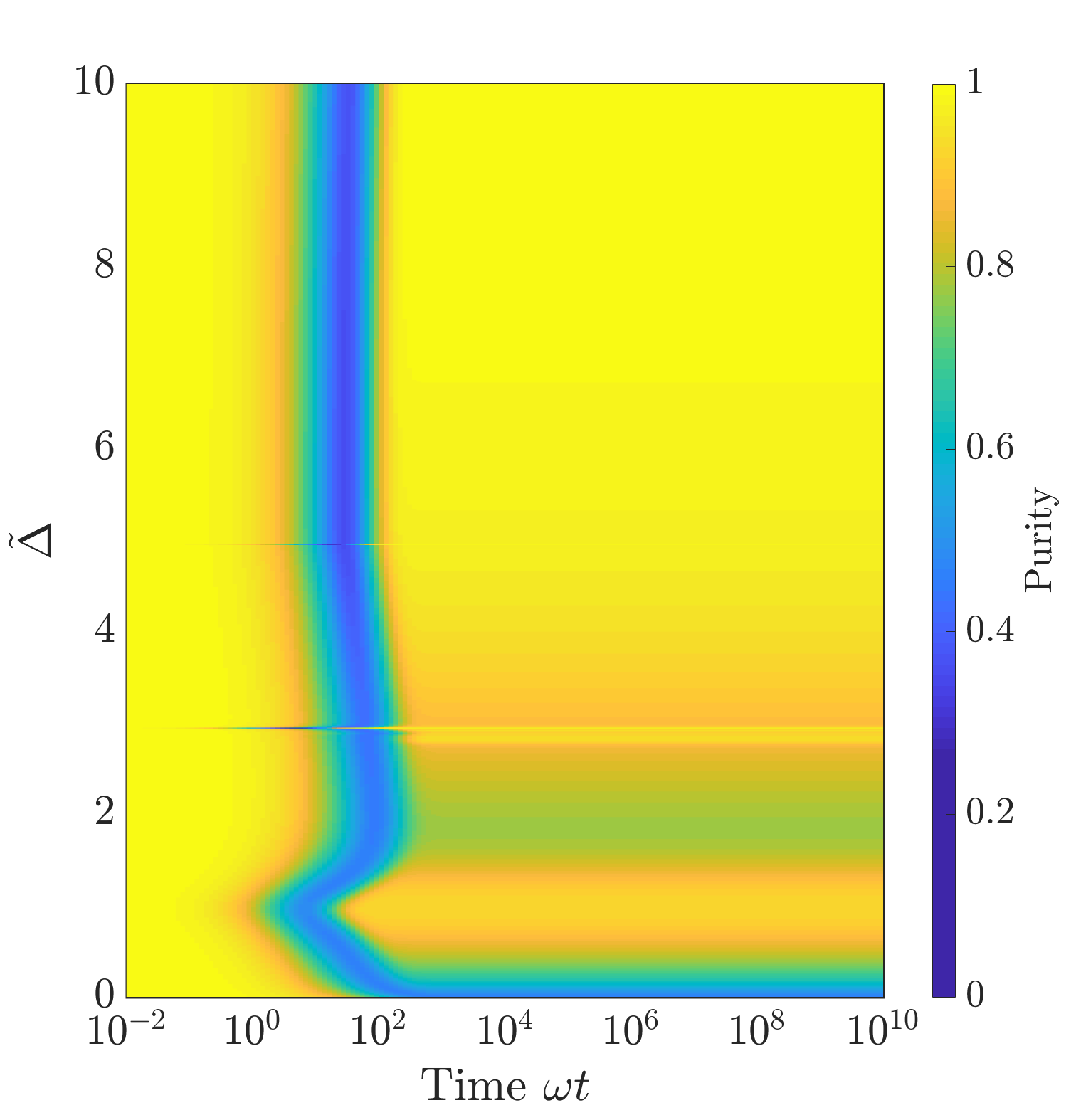}
    \caption{(Top Row) The spin down population dynamics as a function of $\tilde{\Delta}$. (Left) Both baths at zero temperature. (Right) $(T_\mathrm{s}, T_\mathrm{b})/\omega=(2.5,0)$. $\tilde{\lambda}=0.2$. (Bottom Row) The corresponding purity dynamics.}
    \label{fig:RedfieldtDep}
\end{figure}

We now turn to the temporal profile of the Redfield dynamics. As we are interested in the spin flux, it is intuitive to initiate the dynamics with all population in the spin up state. Specifically we choose the initial state to be the product state $|\psi(t=0)\rangle=|\uparrow\rangle\otimes|N_\mathrm{b}=0\rangle$. In Fig.\ref{fig:RedfieldtDep} we show both the case of $(T_\mathrm{s},T_\mathrm{b})=(0,0)$ and $(T_\mathrm{s},T_\mathrm{b})=(2.5~\omega,0)$, corresponding to the blue (zero bias) and the red (strong bias) lines in Fig.~\ref{fig:RedfieldNESS}.  The spin down population as a function of time and $\tilde{\Delta}$ for the two cases is shown in the top row, and the purity as a function of time and $\tilde{\Delta}$ is shown in the bottom row.  First of all, the transient dynamics ($\omega t<10^2$) is near identical regardless of the bath character, as the system Hamiltonian is dominant. This can be confirmed by the corresponding purity dynamics shown in the bottom row of Fig.~\ref{fig:RedfieldtDep}.  While in both cases resonance-induced spin transport sensitivity takes place (the peaks at $\tilde{\Delta}=1,3,5$), only \textit{with finite temperature bias does it persist to the steady state.} This is expected since without bias the steady state solution is guaranteed to be the thermal state $e^{-H_\mathrm{sys}/T}/Z$, which is not sensitive to parameters.

Comparing the population and the purity dynamics it is clear that the state mixing coincides with the emergence of the resonance-induced transport sensitivity. In the case of unbiased system, the purity drops from unity (a pure initial state) to an intermediate value and is eventually restored to unity (a pure final state,  which is the ground state of the system). This picture changes as bias becomes finite, where the purity is never recovered fully and a net flux is maintained at the steady state.  

\newcommand{\noopsort}[1]{} \newcommand{\printfirst}[2]{#1}
  \newcommand{\singleletter}[1]{#1} \newcommand{\switchargs}[2]{#2#1}

\end{document}